\documentclass[usenatbib,onecolumn]{mnras}
\usepackage[T1]{fontenc}
\usepackage{ae,aecompl}

\usepackage[dvipdfmx]{graphicx}	
\usepackage{amsmath}	
\usepackage{amssymb}	
\usepackage{color}
\usepackage{dcolumn}
\usepackage{bm}
\usepackage{comment}
\usepackage{subfigure}

\newcommand{\beq}{\begin{equation}}
\newcommand{\beqa}{\begin{eqnarray}}
\newcommand{\eeq}{\end{equation}}
\newcommand{\eeqa}{\end{eqnarray}}

\newcommand{\simgt}{\lower.5ex\hbox{$\; \buildrel > \over \sim \;$}}
\newcommand{\simlt}{\lower.5ex\hbox{$\; \buildrel < \over \sim \;$}}

\newcommand{\bd}[1]{\mbox{\boldmath $#1$}}

\title[Imprint of $f(R)$ gravity on weak lensing I]{
The imprint of $f(R)$ gravity on weak gravitational lensing I:
Connection between observables and large-scale structure
}

\author[Y.Higuchi et al.]{Yuichi Higuchi$^{1}$\thanks{E-mail: yuichi.higuchi@nao.ac.jp} 
and Masato Shirasaki$^{2}$
\\
$^{1}$Astronomy Data Center, National Astronomical Observatory of Japan, 2-21-1 Osawa, Mitaka, Tokyo 181-8588, Japan\\
$^{2}$Division of Theoretical Astronomy, National Astronomical Observatory of Japan, 2-21-1 Osawa, Mitaka, Tokyo 181-8588, Japan\\
\\
}

\date{Accepted 4/5/2016. Received 3/3/2016; in original form 3/3/2016}

\pubyear{2016}

\begin{document}
\label{firstpage}
\pagerange{\pageref{firstpage}--\pageref{lastpage}}
\maketitle

\begin{abstract}
We study the effect of $f(R)$ gravity on the statistical properties
of various large-scale structures which can be probed 
in weak gravitational lensing measurements.
A set of ray-tracing simulations of gravitational lensing in $f(R)$ gravity 
enables us to explore cosmological information on 
(i) stacking analyses of weak lensing observables 
and (ii) peak statistics in reconstructed lensing mass maps.
For the $f(R)$ model proposed by Hu \& Sawicki,
the measured lensing signals of dark matter haloes in the stacking analysis would show a $\simlt10\%$ difference between
the standard $\Lambda$CDM and the $f(R)$ model 
when the additional degree of freedom in $f(R)$ model would be 
$|f_{\rm R0}|\sim10^{-5}$.
Among various large-scale structures to be studied in stacking analysis, troughs, i.e, underdensity regions 
in projected plane of foreground massive haloes, could be 
promising to constrain the model with $|f_{\rm R0}|\sim10^{-5}$, 
while stacking analysis around voids is found to be difficult to improve the constraint of $|f_{\rm R0}|$ even in future lensing surveys with a sky coverage of $\sim1000$ square degrees.
On the peak statistics, we confirm the correspondence between local maxima and dark matter haloes along the line of sight, 
regardless of the modification of gravity in our simulation.
Thus, the number count of high significance local maxima would 
be useful to probe the mass function of dark matter haloes even in 
the $f(R)$ model with $|f_{\rm R0}|\simlt10^{-5}$.
We also find that including local minima in lensing mass maps would be helpful to improve the constant on $f(R)$ gravity down to $|f_{\rm R0}|=10^{-5}$ in ongoing weak lensing surveys.
\end{abstract}


\section{INTRODUCTION}
The standard cosmological model called $\Lambda$CDM model
has been established by various astronomical observations 
such as measure of distance to supernovae, 
spatial distribution of galaxies and cosmic microwave background 
(e.g., \citealt{1997ApJ...483..565P, 2006PhRvD..74l3507T, 2015arXiv150201589P}). 
Although the $\Lambda$CDM model is 
consistent with observational results 
within the statistical uncertainty, 
the physical origin of accelerating expansion of the Universe 
is still uncertain.
At present, there exist two physical models 
to explain the cosmic acceleration at redshift of $z\simlt1$:
dynamical dark energy model and modified gravity theory.
The former would realize the accelerating expansion of the Universe
within the theory of general relativity by introducing unknown energy,
while the latter does not require an exotic form of energy but modify 
the basic equation of gravitational action.
In order to distinguish these two scenarios,
the measurement of gravitational growth of cosmic matter density 
would be essential 
because the modification of gravity could 
induce some specific features in matter density distribution 
in the Universe.

Vainshtein \citep{1972PhLB...39..393V} 
or chameleon mechanisms 
\citep{2004PhRvD..69d4026K, 2004PhRvL..93q1104K, 2004PhRvD..70l3518B}
are among the most interesting features in modified gravity theory,
which would guarantee the success of general relativity in the solar system.
However, modification of gravity can 
affect the evolution of matter density distribution on extragalactic scales.
For instance, several $N$-body simulations indicate 
the modification of gravity would affect properties of 
group or cluster-sized haloes 
\citep{2011PhRvD..84h4033L, 2014MNRAS.440..833A, 2015MNRAS.449.3635H}.
In particular, the recent simulation in \citet{2015arXiv151101494A} 
showed that matter densities of haloes at central regions 
can increase up to $40\%$ in some cases.
Furthermore, the modification should give prominent effects 
on low density regions such as cosmic voids 
in which screening mechanisms do not work.
Therefore, the differences of properties in low density regions have been investigated in previous numerical simulations
\citep{2012MNRAS.421.3481L, 2015MNRAS.451.4215Z, 2015arXiv151101494A}.
$N$-body simulations of modified gravity showed that
the number count of voids defined in the spatial distribution of haloes
would be about $2$ times smaller than that in general relativity,
while the matter density around voids would show 
a difference with a level of a few percent \citep{2015MNRAS.451.1036C}.
While numerical simulations have been advanced of understanding 
the statistical properties of cosmic matter density in modified gravity,
some of those properties in simulations 
are based on three-dimensional matter density distribution 
and thus cannot be observed 
in actual observations directly.
Therefore, it is important to investigate {\it how the imprint of modified gravity on cosmic matter density would appear in real observables}.
Among various observables, we would focus on 
weak gravitational lensing effect on galaxies in imaging surveys.

Gravitational lensing is an unbiased 
probe of the matter distribution in the Universe. 
Small image distortions of distant galaxies are 
caused by intervening mass distribution. 
Small distortion caused by the large-scale structure of the Universe is
called cosmic shear. 
It contains, in principle, rich information on the matter distribution 
at small and large scales and the evolution over time. 
Image distortion induced by gravitational lensing is, 
however, very small in general. Therefore, we need
statistical analyses of the cosmic shear signal 
by sampling a large number of distant galaxies in order to extract cosmological information from gravitational lensing.  
Ongoing and future galaxy imaging surveys are aimed 
at measuring projected matter distribution 
over several thousand square degrees
and the averaged matter distribution around possible large-scale
structures such as galaxies, galaxy clusters and cosmic voids.
Clearly, we need to understand the relation between 
cosmic matter density and the lensing observables 
when examining the modified gravity with galaxy imaging surveys.

In this paper, we perform ray-tracing simulation of 
gravitational lensing in the modified gravity and
explore the relation between large-scale structures 
and the lensing observables in galaxy imaging surveys 
under the modification of gravity.
In comparison with the recent study in \citet{2015JCAP...10..036T},
we properly take into account the deflection of light ray 
along the line of sight in our gravitational lensing simulations.
We consider the two observables to reveal the matter density distribution 
in the Universe with weak gravitational lensing: 
the stacked lensing signals around various large-scale structures
and the peak statistics on reconstructed mass distribution 
from cosmic shear.
The former observables are related with the average matter density
distribution around dark matter haloes and cosmic voids.
In addition, we also focus on the peak statistics to 
extract cosmological information on the abundance 
of large-scale structures.

This paper is organized as follows. 
In Section~\ref{sec:model}, the cosmological model is described. 
In Section~\ref{sec:WL}, we summarize basics of weak lensing 
and analysis methods used in this paper.
We also explain details of our lensing simulation 
and the definition of large-scale structures in Section~\ref{sec.sim}.
In Section~\ref{sec.result}, we provide results of our lensing analysis 
in numerical simulation of modified gravity and compare the results 
of modified gravity model and the $\Lambda$CDM model in detail.
Conclusions and discussions are summarized in Section~\ref{sec.con}.
\section{COSMOLOGICAL MODEL}\label{sec:model}

There exist various extensions to the standard $\Lambda$CDM model.
Here, we consider competing models called $f(R)$ models,
which represents cosmological models with modified gravity.
This model can explain the observed cosmic acceleration at $z \simlt 1$ 
and satisfy the solar system tests with appropriate parameters.
Throughout this paper, we assume a spatially flat universe and adopt the cosmological parameters which follow the result with Planck satellite \citep{2015arXiv150201589P}: 
matter density $\Omega_{\rm m0}=0.315$, dark energy density 
$\Omega_{\Lambda 0}=0.685$, 
Hubble parameter $h=0.673$ and 
the scalar spectral index $n_s=0.945$.

\subsubsection*{f(R) model}

In $f(R)$ model, 
the Einstein-Hilbert action is modified by a general function of the scalar curvature $R$ \citep{2011PhR...505...59N,Li:2011vk, 2015MNRAS.452.3179S},
\beqa
S_{G} = \int {\rm d}^{4}x \sqrt{-g}\left[\frac{R+f(R)}{16\pi G}\right]. \label{eq:action_fR}
\eeqa
The action with Eq.~(\ref{eq:action_fR}) leads the modified Einstein equation as
\beqa
G_{\mu \nu} + f_{R}R_{\mu \nu}-\left(\frac{f}{2} - \Box f_{R} \right)g_{\mu \nu}
-\nabla_{\mu}\nabla_{\nu}f_{R} = 8\pi G T_{\mu \nu},
\eeqa
where $f_{R} \equiv {\rm d}f/{\rm d}R$, $G_{\mu \nu}\equiv R_{\mu \nu}-1/2g_{\mu \nu}R$ and $\Box\equiv\nabla^\alpha\nabla_\alpha$.
Assuming a Friedmann-Robertson-Walker (FRW) metric, one can determine the time evolution of the Hubble parameter in $f(R)$ model 
as follows:
\beqa
H^{2}-f_{R}\left(H\frac{{\rm d}H}{{\rm d}\ln a}+H^2 \right)
+\frac{f}{6}+H^{2}f_{RR}\frac{{\rm d}R}{{\rm d}\ln a} = \frac{8\pi G}{3}\rho_{m}. \label{eq:H_fR}
\eeqa

One can also consider the evolution of matter 
density perturbations in $f(R)$ model.
For sub-horizon modes $(k \simgt aH)$ in the quasi-static limit
\footnote{\citet{2008PhRvD..77l3515D, 2015JCAP...02..034B} have shown that 
the quasi-static approximation becomes quite reasonable
for models with $|f_{R}| \ll 1$ today. 
}, 
the linear growth of matter density perturbation is determined by \citep[e.g.,][]{2007PhRvD..75f4020B}
\beqa
\frac{{\rm d}^2 g_{+}}{{\rm d} a^2}
+\left(\frac{3}{a}+\frac{1}{H}\frac{{\rm d}H}{{\rm d}a}\right)\frac{{\rm d} g_{+}}{{\rm d} a}
-\frac{3\tilde{\Omega}_{\rm m0}a^{-3}}{\left(H/H_0\right)^2 \left(1+f_{R}\right)}
\frac{1-2Q}{2-3Q}\frac{g_{+}}{a^2} = 0, \label{eq:linear_g_fR}
\eeqa
where $\tilde{\Omega}_{\rm m0}$ is the effective matter density at present time.
We can specify this effective density $\tilde{\Omega}_{\rm m0}$ as
\beqa
H_{f(R)} = H_{0}\sqrt{\tilde{\Omega}_{{\rm m}0}a^{-3}
+\tilde{\Omega}_{\rm DE}
\exp\left[-3\int_{1}^{a} {\rm d}a^{\prime} (1+\tilde{w}_{\rm DE}(a^{\prime}))/a^{\prime}\right]},
\eeqa
where $H_{f(R)}$ is given by Eq.~(\ref{eq:H_fR}).
Note that $\tilde{\Omega}$ and $\tilde{w}_{\rm DE}$ are equivalent to $\Lambda$CDM model for small $|f_{R0}|$.
The function $Q$ in Eq.~(\ref{eq:linear_g_fR}) is given by
\beqa
Q(k, a) = -2\left(\frac{k}{a}\right)^2 \frac{f_{RR}}{1+f_{R}}.
\eeqa
Note that the function of $Q$ induces the non-trivial scale dependence of the linear growth rate $g_{+}(k, a)=D(k, a)/a$ in $f(R)$ model,
while the linear growth rate is exactly a function of $a$ in General Relativity.

In this paper, we will consider the representative example of $f(R)$ models
as proposed in \citet{2007PhRvD..76f4004H} (hereafter denoted as HS model),
\beqa
f(R) = -2\Lambda \frac{R^{n}}{R^{n}+\mu^{2n}},
\eeqa
where $\Lambda$, $\mu$ and $n$ are free parameters in this model.
For $R\gg \mu^2$, 
one can approximate the function of $f(R)$ as follows:
\beqa
f(R) = -2\Lambda -\frac{f_{R0}}{n}\frac{\bar{R}_{0}^{n+1}}{R^n},
\eeqa
where $\bar{R}_{0}$ is defined 
by the present scalar curvature of the background space-time
and $f_{R0} = -2\Lambda \mu^2/\bar{R}_{0}^2 = f_{R}(\bar{R}_{0})$.
In the HS model with $|f_{R0}| \ll 1$, 
the background expansion would be almost equivalent to one in $\Lambda$CDM model.
In practice, for $|f_{R0}| \ll 10^{-2}$, 
geometric tests such as distance measurement with supernovae 
could not distinguish between the HS model and the $\Lambda$CDM model \citep{2012PhRvD..85b4006M}.
Nevertheless, measurements of gravitational growth would be helpful to constrain on HS model due to
the scale dependence of growth rate as shown in Eq.~(\ref{eq:linear_g_fR}).
Furthermore, 
the non-linear gravitational growth in the HS model have been studied with cosmological $N$-body simulations
\citep{2008PhRvD..78l3524O, 2009PhRvD..79h3518S, 2013PhRvD..88j3507H, 2014ApJS..211...23Z}.
Such previous works indicate that 
statistics of galaxy groups or clusters 
provide meaningful information about the modification of gravity.
Therefore, combination of statistics between weak lensing and galaxy clusters are among the interesting probes of the HS model.
In the following, we focus on the case of $n=1$.
 
\section{WEAK LENSING}\label{sec:WL}

\subsection{Basics}
We here summarize basics of weak gravitational lensing effects.
When considering the observed position of a source object as 
$\mbox{\boldmath $\theta$}$ 
and the true position as $\mbox{\boldmath $\beta$}$,
one can characterize the distortion of image of a source object by the following 2D matrix
\beqa
A_{ij} = \frac{\partial \beta^{i}}{\partial \theta^{j}}
           \equiv \left(
\begin{array}{cc}
1-\kappa -\gamma_{1} & -\gamma_{2}-\omega  \\
-\gamma_{2}-\omega & 1-\kappa+\gamma_{1} \\
\end{array}
\right), \label{distortion_tensor}
\eeqa
where $\kappa$ is convergence, $\gamma$ is shear and $\omega$ is rotation.

Let us consider the case of General Relativity. 
One can relate each component of $A_{ij}$ to
the second derivative of the gravitational potential as follows
\citep{2001PhR...340..291B}
\beqa
A_{ij} &=& \delta_{ij} - \Phi_{ij}, \label{eq:Aij} \\
\Phi_{ij}  &=&\frac{2}{c^2}\int _{0}^{\chi}{\rm d}\chi^{\prime} q(\chi,\chi^{\prime}) \partial_{i}\partial_{j}\Phi(\chi^{\prime}), \label{eq:shear_ten}\\	
q(\chi,\chi^{\prime}) &=& \frac{r(\chi-\chi^{\prime})r(\chi^{\prime})}{r(\chi)},
\eeqa
where $\chi$ is the comoving distance and $r(\chi)$ represents the comoving angular diameter distance.
Gravitational potential $\Phi$ can be related to matter density perturbation $\delta$ 
through Poisson equation.
Therefore, convergence can be expressed as the weighted integral of $\delta$ along the line of sight
\beqa
\kappa = \int_{0}^{\chi_{s}}W_{\kappa}(\chi, \chi_{s}) \delta,
\label{eq:kappa_delta}
\eeqa
where 
$\chi_s$ is comoving distance to source galaxies and 
$W_{\kappa}(\chi)$ is the lensing weight function defined as
\beqa
W_{\kappa}(\chi, \chi_{s}) = \frac{3}{2}\left(\frac{H_{0}}{c}\right)^2 \Omega_{\rm m0}\, q(\chi_s, \chi)(1+z(\chi)).
\eeqa
In general, the lensing equation would be governed by so-called lensing potential $(\Phi+\Psi)/2$ where $\Phi$ and $\Psi$ are the Bardeen potentials appeared in metric perturbation in the Newtonian gauge.
The lensing potential in $f(R)$ gravity would be governed by 
the Poisson equation in General relativity, making 
Eqs.~(\ref{eq:Aij}), (\ref{eq:shear_ten}) and 
(\ref{eq:kappa_delta}) available in the HS model with $|f_{R0}| \ll 1$.
Therefore, equations for gravitational lensing in $f(R)$ gravity become the same as in general relativity \citep{2014MNRAS.440..833A}. 

\subsection{Statistical Analysis}
While reduced shear $g_{i}=\gamma_{i}/(1-\kappa)$ 
can be estimated from the ellipticity of galaxy
\citep[e.g.,][]{1995A&A...294..411S, 2001PhR...340..291B}, 
the intrinsic shape would be dominated in 
the measured ellipticity of individual galaxy 
in typical ground-based imaging surveys.
Therefore, we need to perform statistical analyses with a large number of source galaxies.
Here, we introduce two statistical methods to 
study the large-scale structures with weak lensing measurement.

\subsubsection{Stacked lensing}
\label{sec.stlen}
Stacking analysis of weak lensing observables is a powerful method
to study the average matter density distribution 
around an object of interest.

For a given tracer of large-scale structures (LSS),
we define a stacked signal as 
\beqa
\langle \gamma_{+} \rangle (\theta)
= \langle n_{\rm LSS}(\bd{\phi})
\gamma_{+}(\bd{\theta}+\bd{\phi})\rangle,
\eeqa
where $n_{\rm LSS}$ represents the number density of the tracer.
The stacked signal can be related to the surface excess of convergence
as \citep{2001PhR...340..291B}
\beqa
\langle\gamma_{+}\rangle(\theta)
=\bar{\kappa}(\theta)-\langle\kappa\rangle(\theta),
\label{eq.tancon}
\eeqa
where 
$\langle\kappa\rangle$ denotes 
the azimuthal average profile of convergence and 
$\bar{\kappa}(\theta)$ 
is a mean convergence value within a circular aperture 
of radius $\theta$ defined as 
\beqa
\bar{\kappa}(\theta)=\frac{1}{\pi\theta^2}
\int_{\phi<\theta}{\rm d}^2 \bd{\phi}\, 
\kappa(\bd{\phi}).
\eeqa

In order to constrain on the nature of gravity with LSS,
it is essential to study the lower density regions in the Universe.
Since any modified gravity theories should pass the existing 
robust tests of General relativity in the solar system 
or the high density region in the Universe,
an extra fifth force in the modified gravity model 
might be weak but not completely disappeared.
Therefore, the matter density distribution in the Universe 
would be affected by the fifth force in more efficient way at lower density regions
\citep[e.g.,][]{2012MNRAS.421.3481L, 2013MNRAS.431..749C,2015MNRAS.451.1036C}.
Also, the matter fluctuation at linear scales can be a promising target 
to seek for the signature of modification of general relativity
because the linear growth rate would be dependent on the length scale
as shown in Eq.~(\ref{eq:linear_g_fR}).
The linear growth rate can be measured by stacked signals around galaxies or clusters of galaxies at large angular separations
\citep[e.g.,][]{2011PhRvD..83b3008O}.

Hence, we consider three tracers of LSS, 
named voids, troughs, and haloes in the present paper.
Voids are commonly defined by empty regions or extremely low density regions in the Universe.
\citet{2016MNRAS.455.3367G} has recently proposed 
a new tracer of low density regions in the Universe, which 
is called trough. 
Troughs are defined as underdense regions in the projected 
galaxy distributions.
Under the assumption that galaxy would be the biased tracer of matter distribution in a three-dimensional space, 
troughs would be an effective tracer of matter density distributions with 
$\delta<0$.
Haloes are the counterpart of voids, showing the high contrast of matter density with the typical value of $\delta\simgt100$.
We examine whether the stacked signal around voids or troughs (the tracers of underdensity regions) 
can be affected by the extra fifth force in $f(R)$ model.
On the other hand, we study the stacked signals around haloes
at large scales to explore the scale-dependency of linear growth rate.
In the following, we summarize the model of stacked lensing signal 
around haloes, voids and troughs.

\subsubsection*{Haloes}

We here describe the model of the lensing observable 
$\langle \gamma_{+} \rangle$ around dark matter haloes.
Let us suppose that a density profile of a host halo is
described by the universal NFW profile \citep{1997ApJ...490..493N},
\beqa
\rho_{h}(r)=\frac{\rho_s}{(r/r_s)(1+r/r_s)^2}
\label{eq:NFW},
\eeqa
where $\rho_s$ and $r_s$ are a scale density and a scale radius, respectively.
The parameters $\rho_s$ and $r_s$ 
can be essentially convolved into one parameter, 
the concentration $c_{\rm vir}(M,z)$, 
by the use of two halo mass relations; 
namely, 
$M_{\rm vir}=4\pi r^3_{\rm vir} \Delta_{\rm vir}(z) \rho_{\rm crit}(z)/3$, 
where $r_{\rm vir}$ is a virial radius corresponding to the overdensity 
criterion $\Delta_{\rm vir}(z)$
(as shown in, e.g., \citealt{1997ApJ...490..493N}),  
and $M_{\rm vir}= \int dV \, \rho_h (\rho_s,r_s)$ with the integral performed out to $r_{\rm vir}$.
In this paper, we adopt the functional form of the concentration parameter 
in \citet{2014MNRAS.441.3359D},
\beqa
\log_{10} c_{\rm vir}(M, z) = 
0.537+0.488\exp\left(-0.718z^{1.08}\right)
+(-0.097+0.024z)\log_{10}\left(\frac{M}{2\times10^{12}\, h^{-1}M_{\odot}}\right).
\label{eq:cvir_model}
\eeqa

For a given halo sample
with the mass range of $M_{\rm min}<M<M_{\rm max}$
and the redshift range of $z_{\rm min}<z<z_{\rm max}$,
one can find stacked signals around dark matter haloes 
are expressed as
\citep[e.g.,][]{2015MNRAS.451.1418M}
\beqa
\langle \gamma_{+} \rangle (\theta)
= \int\frac{{\rm d}^2\ell}{(2\pi)^{2}} P_{h\kappa}(\ell)J_{2}(\ell\theta),
\eeqa
where $P_{h\kappa}(\ell)$ represents the halo-convergence 
cross power spectrum and 
$J_{2}(x)$ is the second-order Bessel function.
In this paper, we apply the halo-model approach to model $P_{h\kappa}$.
As follows in \citet{2011PhRvD..83b3008O,2016PASJ...68....4S},
the halo-convergence cross power spectrum spectrum is given by
\beqa
P_{h\kappa}(\ell) &=& 
P^{1h}_{h\kappa}(\ell)+P^{2h}_{h\kappa}(\ell), \\
P^{1h}_{h\kappa}(\ell) &=& 
\frac{1}{\bar{n}_{\rm halo}} 
\int_{z_{\rm min}}^{z_{\rm max}} 
{\rm d}z\, 
\frac{{\rm d}^2V}{{\rm d}z{\rm d}\Omega} 
\int_{M_{\rm min}}^{M_{\rm max}} 
{\rm d}M\, \frac{{\rm d}n}{{\rm d}M} \, 
\frac{(1+z)^3}{\bar{\rho}_{m}(z)}\frac{W_{\kappa}(\chi, \chi_{s})}{r(\chi)^2}  
\tilde{\rho}_{h}\left(k=\frac{\ell}{r(\chi)}\Bigg|z(\chi), M\right), 
\label{eq:Phk_1h}
\\
P^{2h}_{h\kappa}(\ell) &=&
\frac{1}{\bar{n}_{\rm halo}}
\int_{z_{\rm min}}^{z_{\rm max}} 
{\rm d}z\, 
\frac{{\rm d}^2V}{{\rm d}z{\rm d}\Omega}
\left[ \int_{M_{\rm min}}^{M_{\rm max}} 
{\rm d}M\, \frac{{\rm d}n}{{\rm d}M} b_{h}(z, M)
\right]
\frac{W_{\kappa}(\chi, \chi_{s})}{r(\chi)^2}
P^{L}_{m}\left(k=\frac{\ell}{r(\chi)}, z(\chi)\right),
\label{eq:Phk_2h}
\eeqa
where
the volume element is expressed as ${\rm d}^2 V/{\rm d}z{\rm d}\Omega = \chi^2/H(z)$ for a spatially flat universe,
$\tilde{\rho}_{h}$ represents the Fourier transform of Eq.~(\ref{eq:NFW}),
${\rm d}n/{\rm d}M$ and $b_{h}$ denote 
the halo mass function and the linear halo bias, respectively.
In Eqs.~(\ref{eq:Phk_1h}) and (\ref{eq:Phk_2h}),
$\bar{n}_{\rm halo}$ represents the average number density of haloes which
is defined by
\beqa
\bar{n}_{\rm halo} = 
\int_{z_{\rm min}}^{z_{\rm max}} 
{\rm d}z\, 
\frac{{\rm d}^2V}{{\rm d}z{\rm d}\Omega}
\int_{M_{\rm min}}^{M_{\rm max}} 
{\rm d}M\, \frac{{\rm d}n}{{\rm d}M}.
\eeqa
In the case of the standard $\Lambda$CDM model, 
we adopt the model of halo mass 
function and linear halo bias with the overdensity of 
$\Delta=200$ developed in \citet{Tinker:2008ff, Tinker:2010my}.
To remain consistent with our calculation, 
we convert the mass of the host halo using the definition of $\Delta$ 
as shown in \cite{2003ApJ...584..702H}.

\subsubsection*{Voids}

We next consider stacked lensing signals around voids.
For the standard $\Lambda$CDM model,
dark matter density profile of voids have been investigated 
in previous studies \citep[e.g.][]{2004MNRAS.350..517S, 2012MNRAS.421..926P}.
Both theoretical and observational studies have indicated 
that a matter density would be almost constant over
an underdense region and there exists 
a very sharp spike called a ridge at the edge of voids.
These features are found in the $N$-body simulation of $f(R)$ gravity 
but the inner profile of voids would be affected by the modification of gravity \citep[e.g.][]{2015MNRAS.451.1036C}.

In order to make a simple void model 
which includes the properties found in the previous works, 
we consider a spherically symmetric void model 
called a double top-hat model \citep{2013MNRAS.432.1021H}.
In this model, mass density of a given void is expressed as
\beqa
\rho_{v}(r)
=
\left\{ 
\begin{array}{ll}
\rho_{1} & (r\leq R_{1}) \\
\rho_{2} & (R_{1}<r\leq R_{2}) \\
0 & (R_{2} < r) \\
\end{array} \right.
,
\eeqa
where $r$ is a distance from the center of void,
$\rho_{1}$ and $\rho_{2}$ are constant in each region.
Assumed that the total mass between the void region $(r\leq R_{1})$
and the ridge region $(R_{1}<r\leq R_{2})$ should be 
compensated each other, the density contrast of void is given by
\beqa
\delta_{v}(r) 
&=& \rho_{v}/\bar{\rho}_{m}-1 \nonumber \\
&=&
\left\{ 
\begin{array}{ll}
\delta_{1} & (r\leq R_{1}) \\
\delta_{1} \left[1-\left(R_{2}/R_{1}\right)^{3}\right]^{-1} & (R_{1}<r\leq R_{2}) \\
0 & (R_{2} < r) \\
\end{array} \right.
,\label{eq:dtop-hat}
\eeqa
where we introduce the parameter of $\delta_{1} = \rho_{1}/\bar{\rho}_{m}-1$ and $\bar{\rho}_{m}$ represents the mean matter density.
The corresponding shear profile from Eq.~(\ref{eq:dtop-hat})
can be calculated analytically as
\beqa
\gamma_{+} (\theta)
&=& 
W_{\kappa}(\chi_{l}, \chi_{s}) \chi_{l} \, \frac{\delta_{1}}{3\theta^{2}}
\Biggl[
\frac{1}{1-\left(\theta_{2}/\theta_{1}\right)^3}
\Biggl\{
\left(\theta_{2}/\theta_{1}\right)^{3}
\left(2\theta_{1}^2+\theta^2\right)
\sqrt{\theta_{1}^2-\theta^2}
-\left(2\theta_{2}^2+\theta^2\right)
\sqrt{\theta_{2}^2-\theta^{2}}
\Biggr\}\Theta(\theta_{1}-\theta) \nonumber \\ 
&&-\frac{\left(2\theta_{2}^2+\theta^2\right)\sqrt{\theta_{2}^2-\theta^{2}}}{1-\left(\theta_{2}/\theta_{1}\right)^3}
\Theta(\theta-\theta_{1})\Theta(\theta_{2}-\theta)
\Biggr],
\label{eq:shear_void}
\eeqa
where $\chi_{l}$ is the comoving distance to void,
$\theta_{i}$ is defined by $R_{i}/\chi_{l}$
and $\Theta(x)$ represents the Heaviside step function.

\subsubsection*{Troughs}

Troughs have been recently proposed in \citet{2016MNRAS.455.3367G}
as tracers of underdense regions in the Universe.
In this paper,
we define the trough with a given dark matter halo catalog.
Let us consider a halo catalog with the selection function 
of $W_{\rm halo}(z, M)$.
For a given $W_{\rm halo}$, 
the three-dimensional number density of haloes is defined by 
\beqa
n_{\rm halo, 3D}({\bd x})
=\sum_{i}^{N_{\rm halo}}\delta^{(3)}({\bd x}-{\bd x}_{i})
W_{\rm halo}(z_{i}, M_{i}),
\eeqa
where $N_{\rm halo}$ represents the total number of haloes 
in the field of view.
In this paper, we simply consider the functional form of 
$W_{\rm halo}(z, M)$ as
\beqa
W_{\rm halo}\left(z, M\right)=
\left\{
\begin{array}{c}
1\\
\\
0
\end{array}
\begin{array}{c}
{\rm for}\hspace{0.25cm} z_{\rm min}\leq z\leq {\rm z}_{\rm max} \hspace{0.25cm}{\rm and}\hspace{0.25cm} M \geq M_{\rm T}\\
\\
{\rm others}
\end{array}
\right.
,
\eeqa
where $M_{\rm T}$ is a selection criterion for halo mass.
From the halo catalog, we can define the smoothed, projected 
number density of haloes as
\beqa
G\left(\bd{\theta}\right)=
\sum^{N_{\rm halo}}_{i=1}W_{\rm T}\left(|\bd{\theta}-\bd{\theta}_i|\right)
W_{\rm halo}\left(z_i, M_i\right),
\label{eq:num_halo_cyl}
\eeqa
where $W_{\rm T}(x)$ is the weighted function to construct the smoothed density field and set to be the top-hat function as
\beqa
W_{\rm T}\left(|\bd{\theta}-\bd{\theta}_i|\right)=
\left\{
\begin{array}{c}
1/\pi\theta_{\rm T}^2\\
\\
0
\end{array}
\begin{array}{c}
{\rm for}\hspace{0.25cm} |\bd{\theta}-\bd{\theta}_i|<\theta_{\rm T}\\
\\
{\rm others}
\end{array}
\right.
,
\eeqa
where $\theta_{\rm T}$ represents the radius of the filter function.
In the following, we call $\theta_{\rm T}$ as trough radius.
The positions of troughs are then selected by 
the points below the $\alpha$-th percentile $G_{\alpha}$ of the distribution
of $G(\bd \theta)$.

As shown in the above selection of troughs, the statistical property of troughs would be governed by the new random field as
\beqa
\delta_{\rm T}(\bd \theta) = 
\int {\rm d}^2\theta^{\prime} 
W_{\rm T}(|{\bd \theta}-{\bd \theta}^{\prime}|) 
\delta_{\Sigma, h}({\bd \theta}^{\prime}),
\eeqa
where $\delta_{\Sigma, h}$ represents the contrast of 
the projected number density of haloes.
Assumed that the probability $P$ of 
finding $N$ haloes at a given position of ${\bd \theta}$
can be determined by the value of $\delta_{\rm T}(\bd \theta)$,
we can identify the expectation value of $\delta_{\rm T}$ 
for a given $N$ as
\beqa
\langle \delta_{\rm T} | N \rangle = \int_{-1}^{\infty} 
{\rm d}\delta_{\rm T}\, p(\delta_{\rm T}|N),
\eeqa
where $p(\delta_{T}|N)$ represents the conditional probability distribution function of $\delta_{\rm T}$ for a given $N$.
Note that the Bayes' theorem tells 
\beqa
p(\delta_{\rm T}|N) &=& \frac{P(N|\delta_{\rm T})P(\delta_{\rm T})}{P(N)}, \\
P(N) &=& \int_{-1}^{\infty} {\rm d}\delta_{\rm T}\, 
P(N|\delta_{\rm T})P(\delta_{\rm T}).
\eeqa
Therefore, the maximum number of haloes within trough radius 
$N_{\rm max}$ should be determined by 
\beqa
\sum_{N=0}^{N_{\rm max}} P(N) = \alpha,
\eeqa
for a given $\alpha$ to define the position of trough.

\citet{2016MNRAS.455.3367G} proposed a simple model of 
the stacked signal around troughs under the following assumptions:
\begin{itemize}
\item The three-dimensional halo density field 
$n_{\rm halo, 3D}$ can be described as 
a deterministic, biased tracer of the matter density field.
This means that 
the projected matter density contrast $\delta_{\Sigma}$ within the redshift range of the halo catalog is equivalent to $\delta_{\Sigma, h}$.
\item The random field $\delta_{\rm T}(\bd \theta)$ and the convergence field $\kappa(\bd \theta)$ follow a Gaussian distribution. 
\item The probability of $P(N|\delta_{\rm T})$ is given 
by a biased Poisson process with the halo bias of $\bar{b}$.
\end{itemize}
Under these assumptions, 
we can calculate the azimuthally averaged convergence profile around troughs.
Let $K_{i}$ to be the average of convergence field 
over a distance of $\theta =[\theta_{i},\theta_{i+1}]$
from the trough center, i.e.,
\beqa
K_{i} = \frac{1}{\pi(\theta_{i+1}^2-\theta_{i}^2)}
\int_{A_{i}} {\rm d}^2 \theta^{\prime} 
\kappa({\bd \theta}^{\prime}),
\eeqa
where $A_{i}$ represents the $i$-th annuli around the trough.
The average $K_{i}$ over the position of troughs is then given by
\beqa
\langle K_{i}|N_{\rm max} \rangle
= \frac{{\rm Cov}(K_{i}, \delta_{T})}{\sigma_{T}^2}
\frac{\sum_{N=0}^{N_{\rm max}}P(N) \langle \delta_{T}|N \rangle}
{\sum_{N=0}^{N_{\rm max}}P(N)},
\label{eq:kappa_trough}
\eeqa
where 
$\sigma_{T}^2$ is the variance of $\delta_{T}$ and
${\rm Cov}(K_{i}, \delta_{T})$ is the covariance between $K_{i}$ and $\delta_{T}$.
We refer the reader to \citet{2016MNRAS.455.3367G} 
for the function form of $P(N)$ and the derivation of 
${\rm Cov}(K_{i}, \delta_{T})$ and $\sigma_{T}^2$.
Hence, one can calculate the stacked signal around troughs
by using Eqs.~(\ref{eq.tancon}) and (\ref{eq:kappa_trough}).

\subsubsection{Peak Statistics}
\label{subsec:peak}
\subsubsection*{Reconstruction of smoothed convergence}
In addition to stacked lensing,
weak lensing provides a physical method 
to reconstruct the projected matter density field.
The reconstruction is commonly based on 
the smoothed map of cosmic shear.
Let us first define the smoothed convergence map as
\beqa
{\cal K}(\bd{\theta}) = 
\int {\rm d}^2 \bd{\phi}\, \kappa(\bd{\theta}-\bd{\phi}) 
U (\bd{\phi}),
\eeqa
where $U$ is the filter function to be specified below.
We can calculate the same quantity by smoothing the shear field $\gamma$ as
\beqa
{\cal K} (\bd{\theta}) = \int {\rm d}^2 \phi \ \gamma_{+}(\bd{\phi}:\bd{\theta}) Q_{+}(\bd{\phi}), \label{eq:ksm}
\eeqa
where $\gamma_{+}$ is the tangential component of the shear at position $\bd{\phi}$ relative to the point $\bd{\theta}$.
The filter function for the shear field $Q_{+}$ is related to $U$ by
\beqa
Q_{+}(\theta) = \int_{0}^{\theta} {\rm d}\theta^{\prime} \ \theta^{\prime} U(\theta^{\prime}) - U(\theta).
\label{eq:U_Q_fil}
\eeqa
We consider $Q_{+}$ to be defined with a finite extent.
In this case, one finds 
\beqa
U(\theta) = 2\int_{\theta}^{\theta_{o}} {\rm d}\theta^{\prime} \ \frac{Q_{+}(\theta^{\prime})}{\theta^{\prime}} - Q_{+}(\theta),
\eeqa
where $\theta_{o}$ is the outer boundary of the filter function.

In the following, we consider the truncated Gaussian filter (for $U$) as
\beqa
U(\theta) &=& \frac{1}{\pi \theta_{G}^{2}} \exp \left( -\frac{\theta^2}{\theta_{G}^2} \right)
-\frac{1}{\pi \theta_{o}^2}\left[ 1-\exp \left(-\frac{\theta_{o}^2}{\theta_{G}^2} \right) \right], \\
Q_{+}(\theta) &=& \frac{1}{\pi \theta^{2}}\left[ 1-\left(1+\frac{\theta^2}{\theta_{G}^2}\right)\exp\left(-\frac{\theta^2}{\theta_{G}^2}\right)\right],
\label{eq:filter_gamma}
\eeqa
for $\theta \leq \theta_{o}$ and $U = Q_{+} = 0$ elsewhere.
Throughout this paper, we adopt $\theta_{G} = 1$ arcmin 
and $\theta_{o} = 10$ arcmin.
Note that this choice of $\theta_{G}$ is considered to be
an optimal smoothing scale for the detection of massive galaxy clusters
using weak lensing for $z_{\rm source}$ = 1.0 
\citep{2004MNRAS.350..893H}.

The local maxima or minima found in a smoothed convergence map
would have cosmological information originated from 
massive dark matter haloes, voids, 
and the superposition of large scale structures
\citep[e.g.,][]{2004MNRAS.350..893H, 2010MNRAS.402.1049D, 2010PhRvD..81d3519K, 2011PhRvD..84d3529Y, 2016PASJ...68....4S}.
In this paper, we define peaks in the convergence map 
by finding the local maxima or minima.
Peak height is in practice normalized as
$\nu(\bd{\theta})={\cal K}(\bd{\theta})/\sigma_{\rm shape}$
where $\sigma_{\rm shape}$ is the noise variance coming from intrinsic ellipticity of galaxies.
We estimate $\sigma_{\rm shape}$ as follows
\beqa
\sigma_{\rm shape}^2=\frac{\sigma_{\rm e}^2}{2n_{\rm gal}}\int_0^{\theta_{\rm out}}{\rm d}\theta \, Q_{+}^2\left(\theta\right),
\label{eq.noise}
\eeqa
where $\sigma_{\rm e}$ is the rms value of intrinsic ellipticity of galaxies and $n_{\rm gal}$ is the number density of galaxies. 
We assume $\sigma_e=0.4$ and $n_{\rm gal}=30$ arcmin$^{-2}$ which are typical values for ground-based imaging surveys.

\subsubsection*{High significance local maxima and their abundance}

On a smoothed lensing map, 
local maxima with high signal-to-noise ratio 
would be likely caused by galaxy clusters
\citep{2004MNRAS.350..893H}.
We thus locate high-$\nu$ local maxima 
on a convergence map and associate each of them 
with an isolated massive halo along the line of sight.
Supposed that the universal NFW density profile 
(Eq.~\ref{eq:NFW}),
we can calculate the convergence profile $\kappa_{h}$ 
for a given halo analytically \citep{2004MNRAS.350..893H}.

In order to predict peak heights in a convergence map, 
we adopt the simple assumption that each peak position is exactly at the halo center.
Under this assumption, 
the peak height in absence of shape noise is given by
\beqa
{\cal K}_{{\rm peak}, h} = \int {\rm d}^2\phi \, U(\phi; \theta_{G},\theta_{o})\kappa_{h}(\phi).
\label{eq:kpeak_nfw}
\eeqa
The actual peak height on a noisy convergence map 
is determined not by Eq.~(\ref{eq:kpeak_nfw}), 
but by a probability distribution function \citep{2010ApJ...719.1408F}.
The probability distribution function of the measured peak height ${\cal K}_{\rm peak, obs}$
with a given ${\cal K}_{{\rm peak}, h}$ is denoted
by ${\rm Prob}({\cal K}_{\rm peak, obs}|{\cal K}_{{\rm peak}, h})$ 
in this paper.
The detailed functional form of ${\rm Prob}({\cal K}_{\rm peak, obs}|{\cal K}_{{\rm peak}, h})$
is found in \citet{2015MNRAS.453.3043S}.

We identify local maxima in a smoothed lensing map and 
match each peak with a massive dark matter halo along the line of sight.
The simple peak count is useful 
to extract the information of the abundance of massive clusters. 
One can select the lensing peaks by its peak height.
For a given threshold of peak height $\nu_{\rm thre}$, 
one can predict the surface number density of peaks
with $\nu > \nu_{\rm thre}$ as follows
\citep{2004MNRAS.350..893H}:
\beqa
N_{\rm peak}(>\nu_{\rm thre}) 
=
\int {\rm d}z\, {\rm d}M\, 
\frac{{\rm d}^2V}{{\rm d}z{\rm d}\Omega} \frac{{\rm d}n}{{\rm d} M}\, S(z, M|\nu_{\rm thre}). \label{eq:npeak}
\eeqa
where $S(z, M|\nu_{\rm thre})$ expresses the selection function
of weak lensing selected clusters
for a given threshold of $\nu_{\rm thre}$. 
It is given by
\beqa
S(z, M|\nu_{\rm thre}) 
=
\int_{\nu_{\rm thre}\sigma_{\rm shape}}^{\infty} 
{\rm d}{\cal K}_{\rm peak, obs}\, 
{\rm Prob}({\cal K}_{\rm peak, obs}|\, {\cal K}_{{\rm peak},h}(z, M)).
\eeqa

\section{SIMULATION}\label{sec.sim}

In order to study the relation of weak lensing statistics with the underlying large-scale structures, we perform the ray-tracing simulation of gravitational lensing with a set of $N$-body simulations.

\subsection{$N$-body and Ray-tracing simulations}
We first run a $N$-body simulation to generate
a three-dimensional matter density field. 
We use the adaptive mesh refinement code {\tt ECOSMOG}
for a wide class of modified gravity \citep{Li:2011vk}. 
The simulation volume has a comoving box length of $240\, h^{-1}$Mpc, resolved using a uniform $512^3$ root grid and 7 levels of 
mesh refinement, implying a maximum comoving spatial resolution of $3.6\, h^{-1}$kpc.
We proceed the mesh refinement when 
the effective particle number in a grid cell would be larger than 8.
The density assignment and force interpolation in a cell is performed
by the triangular shaped cloud (TSC) method.
We generate the initial conditions using a parallel code 
{\tt mpgrafic} developed by \citet{Prunet:2008fv}.
The initial redshift is set to $z_{\rm init}=85$, 
where we compute the linear matter transfer function using 
{\tt linger} \citep{1995astro.ph..6070B}.
In simulations, we adopt the following cosmological parameters : 
matter density $\Omega_{\rm m0}=0.315$, dark energy density 
$\Omega_{\Lambda 0}=0.685$, 
Hubble parameter $h=0.673$ and 
the scalar spectral index $n_s=0.945$.
For the HS model, 
we consider two models with $|f_{\rm R0}|=10^{-5}$ and $10^{-6}$,
referred as F5 and F6, respectively.
Note that we can safely set $\tilde{w}_{\rm DE}=-1$ and 
$\nabla (\Psi + \Phi)/2 = 4\pi G {\bar \rho}_{m} \delta a^{2}$ 
in the model with $|f_{\rm R0}| \ll 1$. 

\begin{figure}
\centering
\includegraphics[width=0.28\columnwidth]{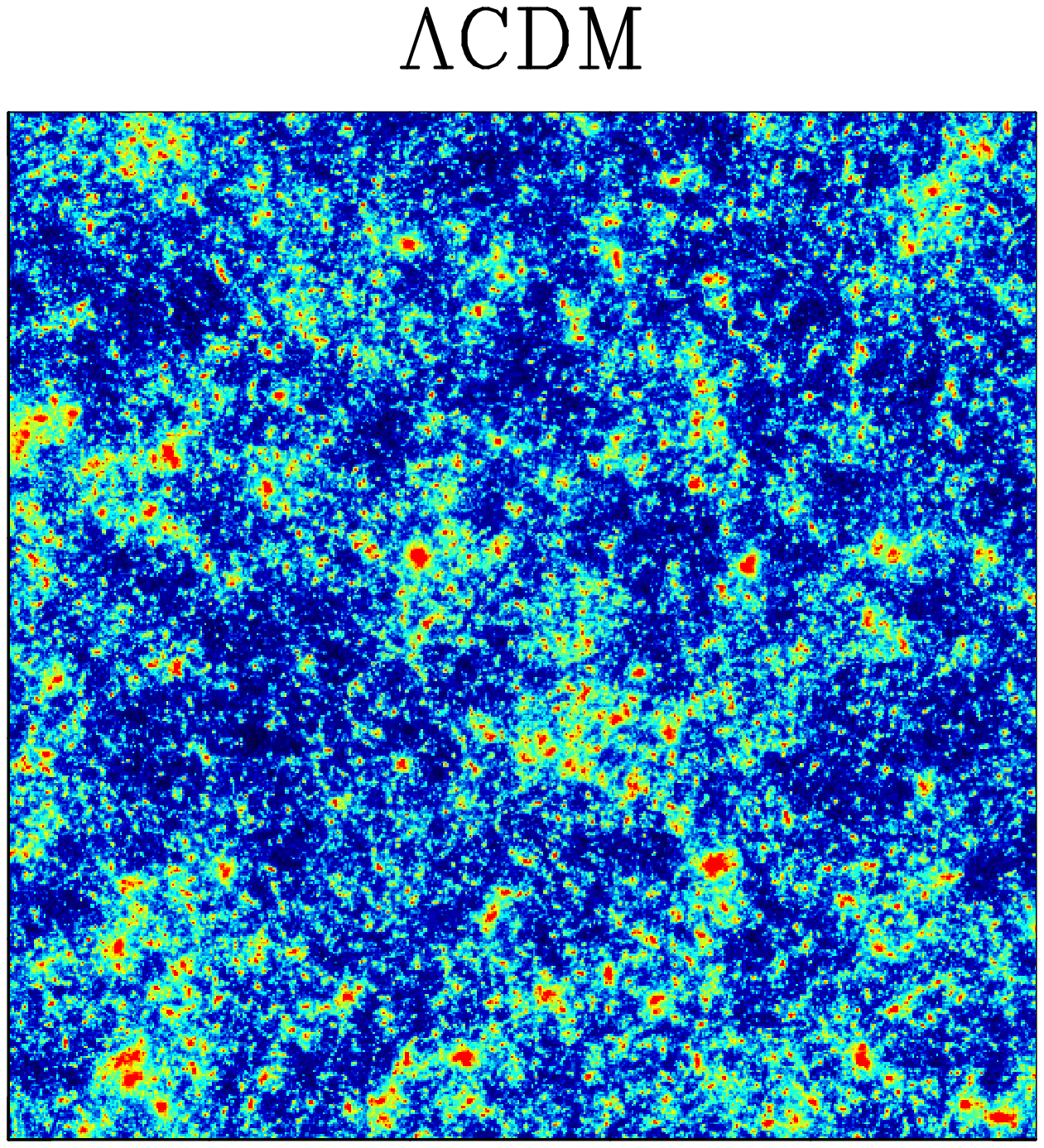}
\includegraphics[width=0.28\columnwidth]{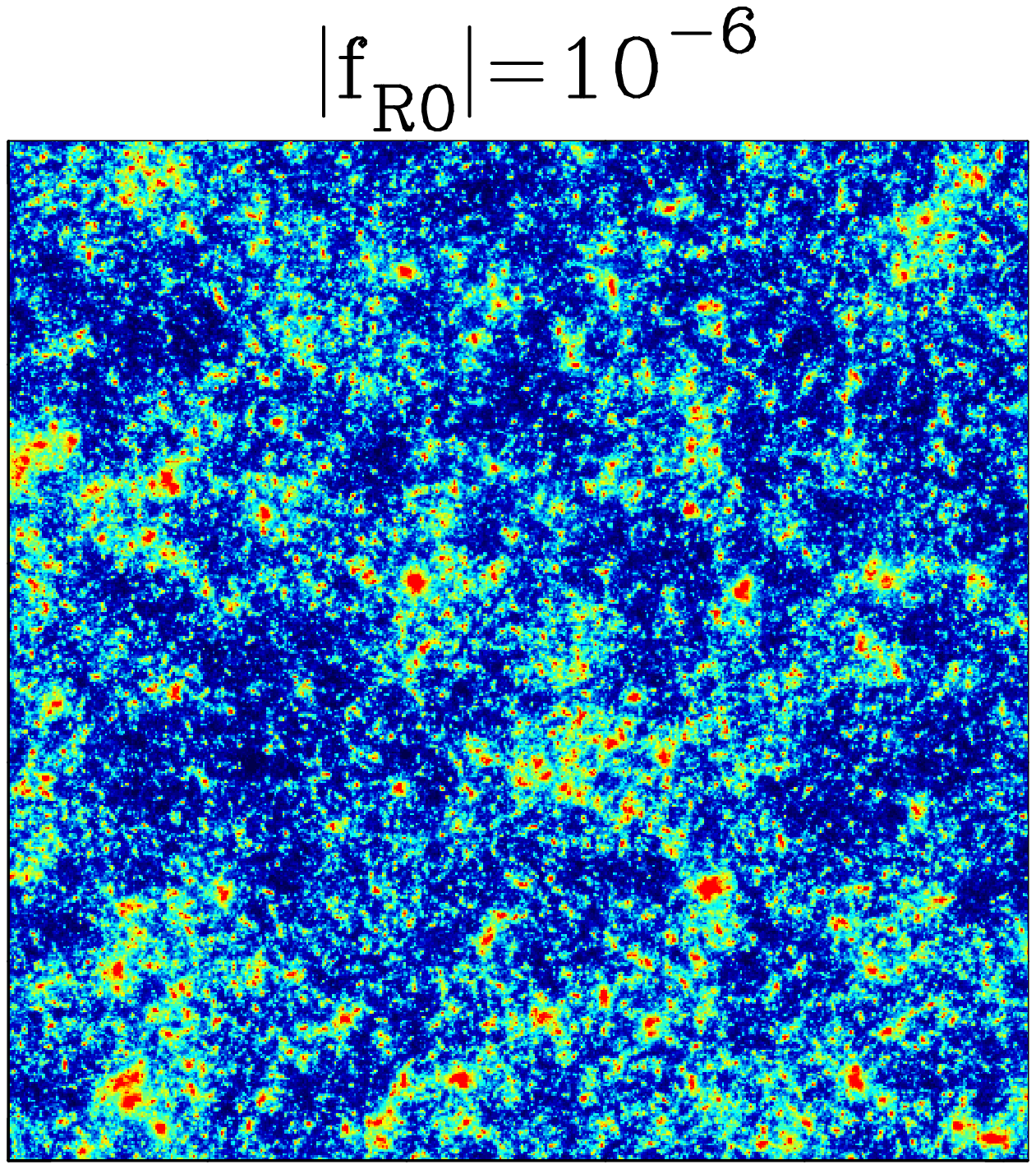}
\includegraphics[width=0.28\columnwidth]{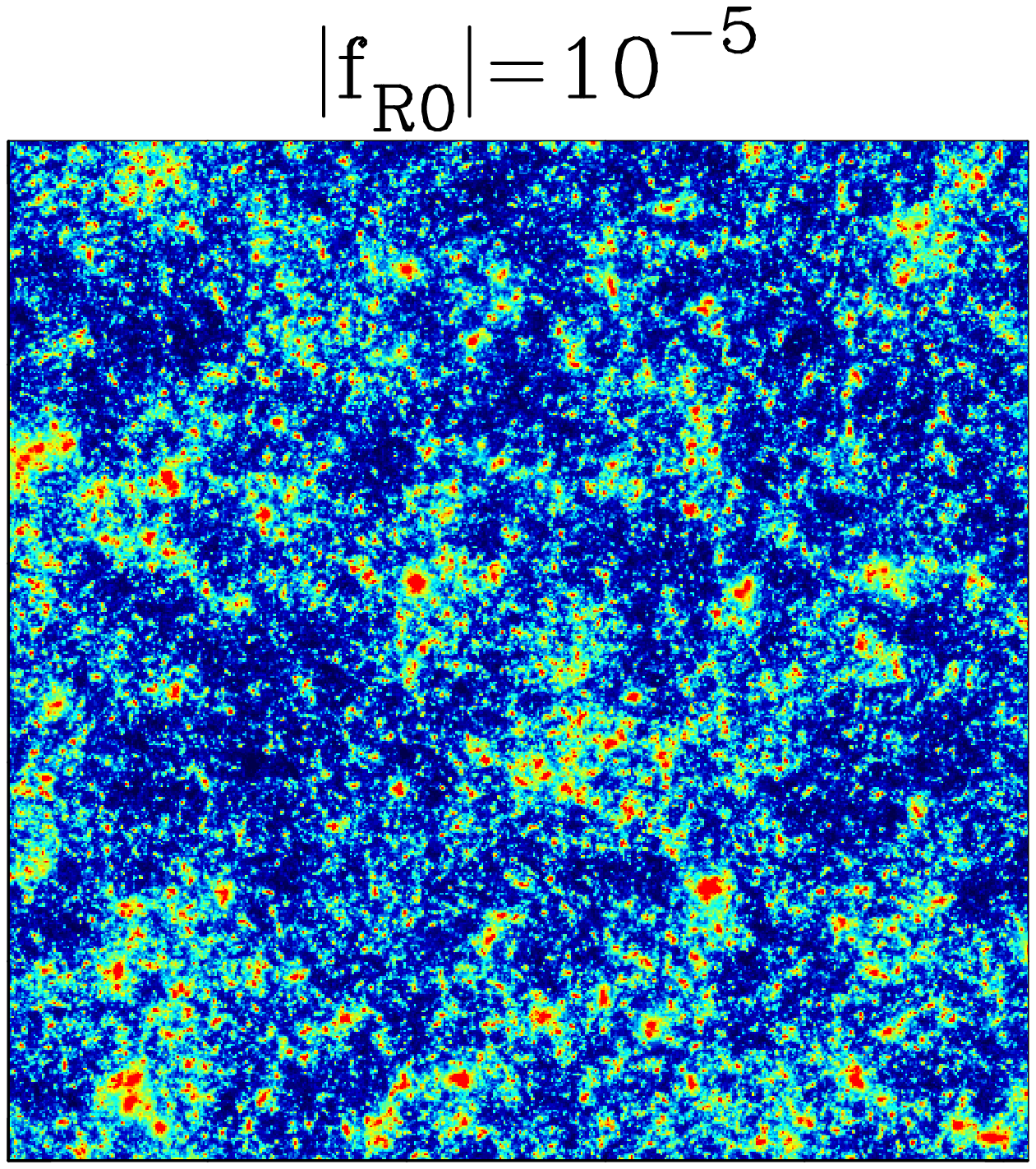}
\caption{
	Convergence map for three different cosmological models.
	The source redshift is set to be unity.
	Each map covers $5\times5$ square degrees.
	It is difficult to distinguish the differences among three maps
	by eyes.
	}
\label{fig:kappa_map}
\end{figure} 

\begin{figure}
\begin{center}
\includegraphics[width=0.45\columnwidth]{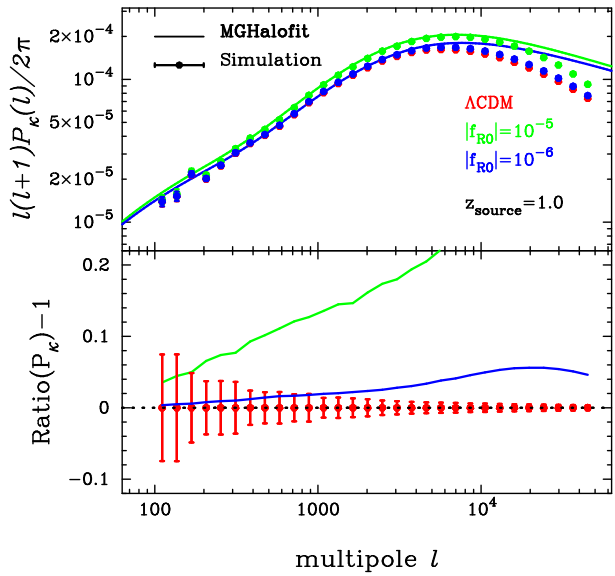}
\end{center}
\caption{Convergence power spectrum for three different models.
In the top panel, the coloured points with error bar show 
the average power spectrum over 100 realizations of lensing maps.
The coloured solid line represents the theoretical model 
of power spectrum with the fitting formula of the non-linear
matter power spectrum \citep{2014ApJS..211...23Z}.
In the bottom panel, we show the difference between the HS model 	and the fiducial $\Lambda {\rm CDM}$ model obtained 
from our simulation.
In both panels, the error bar corresponds to the standard deviation 
of the mean 	estimated from 100 maps.}
\label{fig:Pkappa}
\end{figure}

For ray-tracing simulations of gravitational lensing, 
we generate light-cone outputs using multiple simulation boxes
in the following manner. 
Our simulations are placed to cover the past light-cone of a hypothetical observer with an angular extent $5^{\circ}\times 5^{\circ}$, 
from $z=0$ to 1, similar to the methods in  
\citet{2000ApJ...537....1W},\citet{2001MNRAS.327..169H},
and \citet{2009ApJ...701..945S}.
Details of the configuration are found in the last reference.
The angular grid size of our maps is 
$5^{\circ}/4096\sim 0.075$ arcmin.
For a given cosmological model,
we first store particle data of $N$-body simulation at various redshifts.
We then randomly shift the simulation boxes
in order to avoid the same structure appearing 
multiple times along a line-of-sight.
In total, we generate 100 independent lensing maps 
with the source redshift of $z_{\rm source}=1$
from our $N$-body simulation.
Figure~\ref{fig:kappa_map} shows an example of our simulated convergence map for different cosmological models.
Although it is difficult to distinguish with three models by eyes,
the lensing statistics would enable us to clarify the differences among these.
In Figure~\ref{fig:Pkappa}, we compare the convergence power spectrum for three cosmological models.
The points with error bar represent the simulation results 
for the $\Lambda$CDM model and the HS model with $|f_{\rm R0}|=10^{-5}$ and $10^{-6}$.
The green and blue solid lines correspond to the theoretical model 
of convergence power spectrum for the case of $|f_{\rm R0}|=10^{-5}$ and $10^{-6}$, respectively.
Under the Limber approximation \citep{Limber:1954zz,Kaiser:1991qi}
and Eq.~(\ref{eq:kappa_delta}), 
one can calculate the convergence power spectrum as 
\beqa
P_{\kappa}(\ell) &=& \int_{0}^{\chi_s} {\rm d}\chi \frac{W_{\kappa}(\chi, \chi_s)^2}{r(\chi)^2} 
P_{\delta}\left(k=\frac{\ell}{r(\chi)},z(\chi)\right)
\label{eq:kappa_power},
\eeqa
where $P_{\delta}(k)$ is the three dimensional matter power spectrum.
For the non-linear power spectrum of matter density $P_{\delta}$,
we adopt the fitting formula called {\tt MGHalofit} 
\citep{2014ApJS..211...23Z}.
According to Figure~\ref{fig:Pkappa}, 
our lensing simulation is found to be consistent with previous modified gravity simulations.

\subsection{Definition of Large-Scale Structures}

\subsubsection*{haloes}
In each output of the $N$-body simulation, 
we locate dark matter haloes using 
the standard friend-of-friend (FOF) algorithm
with the linking parameter of $b=0.2$.
We define the mass of each halo by the spherical overdensity mass with $\Delta=200$, which is denoted by $M_{200}$.
The position of each halo is defined by 
the position of the particle located at the potential maximum
in each FOF group.
In the following analysis, 
we use haloes with a mass $M_{200}$ 
greater than $10^{13}h^{-1}{\rm M}_{\odot}$.
Using the FOF haloes, we construct mock group catalogs 
on the light cone 
by arranging the simulation outputs in the same manner 
as the ray-tracing simulation.
We mark the positions of the haloes in the lensing map.
In summary, our mock catalogs contain data 
about the masses, redshifts, and angular positions on
the lensing map for the haloes.

\subsubsection*{Voids}
In order to define void in each realization, 
we employ the public code {\tt Void Finder} \citep{2009ApJ...699.1252F}
on the void finding algorithm 
developed by \citet{2004ApJ...607..751H} and 
\citet{2005ApJ...620..618H}. 
It finds small number density regions of haloes and make spheres. 
Then, radius of those spheres are enlarged and spheres are marginalized with marginalization criteria.
We adopt the same parameters found in \citet{2009ApJ...699.1252F} except for the minimum radius $\xi$. 
We use $\xi=10$ Mpc as the minimum radius of voids.
The number count of voids does not strongly depend on this value for voids whose radius are more than $15$ Mpc \citep{2013MNRAS.432.1021H}.  
When running the {\tt Void Finder}, 
we use haloes at $0.15\leq z\leq 0.65$ with masses 
larger than $M_{200}\ge10^{13}\, h^{-1}M_{\odot}$. 
In order to select legitimate voids, 
we select voids whose centers are more than 
the effective radius of voids away from the edges of 
simulation boxes. 

\subsubsection*{Troughs}
In this paper, we define trough as centers
of cylindrical regions with low number density of haloes as summarized 
in Section~\ref{sec.stlen}.
In order to define positions of troughs, 
we count the number count in each cylinder $G(\bd \theta)$
and then select the set of trough positions 
as the points below the 20th percentile
of the distribution of $G(\bd{\theta})$.
In our study, we estimate the number count 
for the circles spaced by $0.75$ arcmin.
For the selection, we set $z_{\rm min}=0.2$ and $z_{\rm high}=0.6$,
while we examine the three mass threshold cases for $M_{\rm T}=$
$10^{13}h^{-1}M_\odot$, 
$5\times10^{13}h^{-1}M_\odot$
and
$10^{14}h^{-1}M_\odot$.
Furthermore, we study the dependence of the trough radius 
on our results by considering 
$\theta_{\rm T}=5$, $10$, $20$ and $30$ arcmin. 

\section{RESULTS}
\label{sec.result}

In the following, we summarize the results on our lensing analyses
with 100 ray-tracing simulations.

\subsection{Stacked Lensing}
\subsubsection{haloes}
\label{sec.haloestack}

We perform the stacked analysis around haloes
by selecting their masses and redshifts.
We consider the three redshift bins of 
$0.1\leq z \leq0.3$, $0.3\leq z \leq0.5$, 
and 
$0.5\leq z \leq0.7$ and 
three mass bins of 
$10^{13} h^{-1}M_{\odot}\leq M_{200} < 10^{13.5}h^{-1}M_{\odot}$,
$10^{13.5} h^{-1}M_{\odot}\leq M_{200} < 10^{14}h^{-1}M_{\odot}$
and 
$10^{14}h^{-1}M_{\odot} \leq M_{200} < 10^{15}h^{-1}M_{\odot}$. 
In the stacking analysis, 
we measure the azimuthally averaged, logarithmically spaced radial profile
of tangential shear in the radial range of $\theta = 1-100\, {\rm arcmin}$ 
around the center of each cluster and then stack the shear profiles over the haloes found in each realization.
The typical number of haloes in the stacking analysis ranges 
from $40$ to $2000$ in each realization.

\begin{figure}
\centering
\includegraphics[width=0.47\columnwidth]{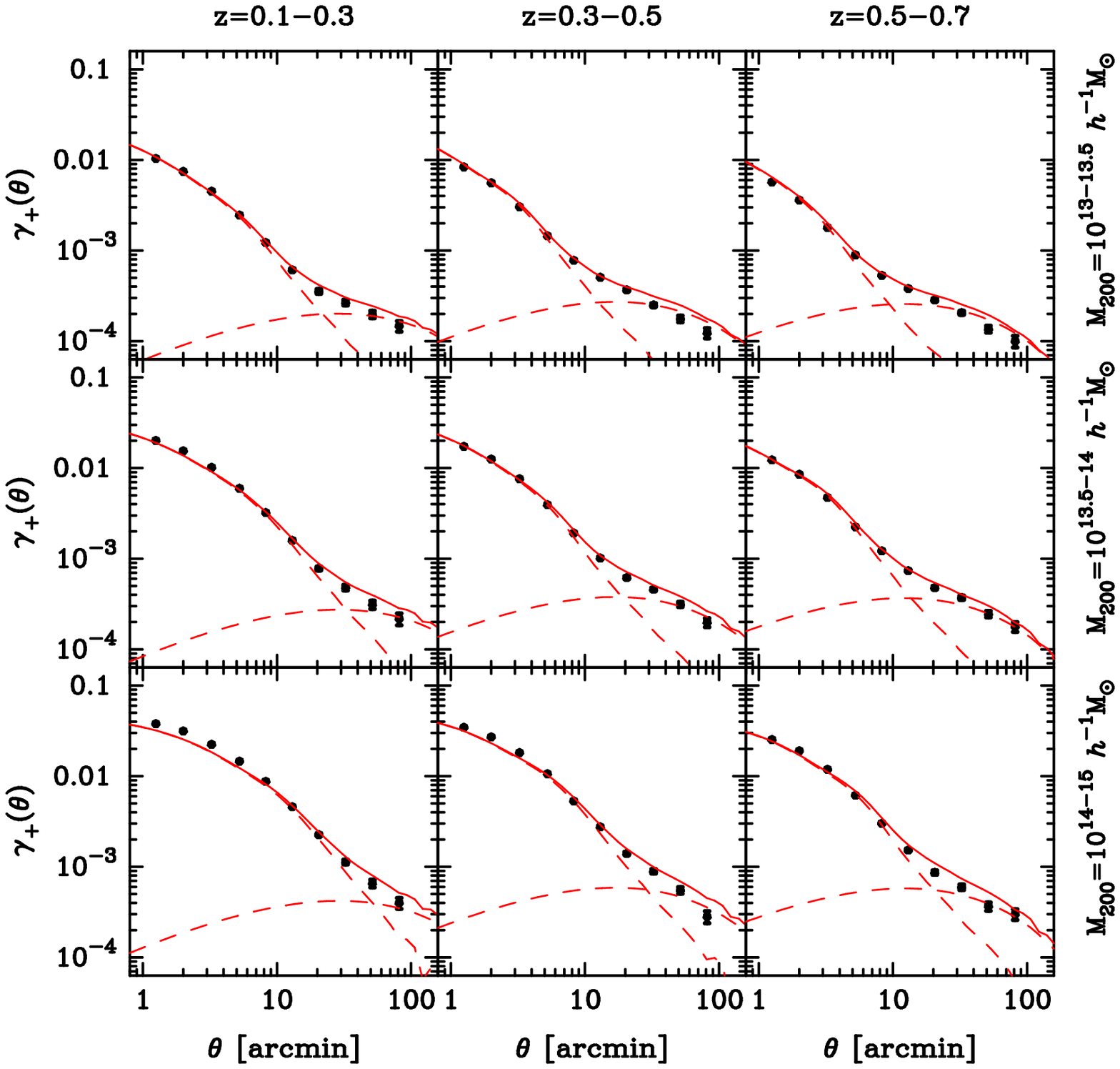}
\includegraphics[width=0.47\columnwidth]{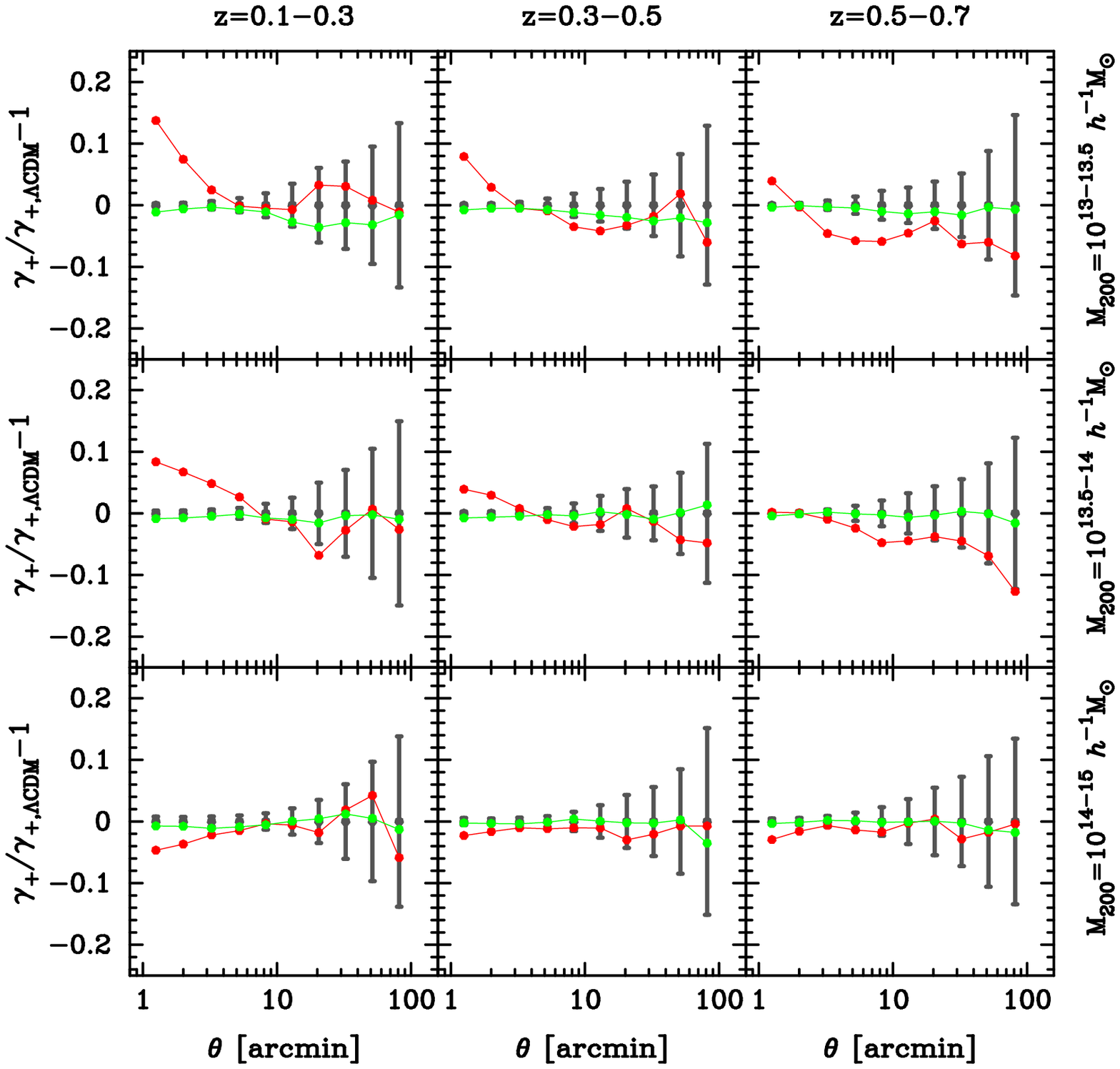}
\caption{
	Stacked profile of haloes.
	We divide the haloes by their halo masses and redshifts.
	{\it Left}: The result for the standard $\Lambda$CDM model.
	The black point with error bar shows the result obtained from
	100 ray-tracing simulations, while the red solid line shows the corresponding halo-model prediction as shown in Section~\ref{sec.stlen}.
	The two red dashed lines are the one-halo and two-halo terms, respectively.
	{\it Right}: 
	The comparison of results with $f(R)$ gravity and $\Lambda$CDM.
	The each panel shows the ratio of stacked signal between the case of 
	$f(R)$ gravity and $\Lambda$CDM model.
	The red line shows the case of $|f_{\rm R0}|=10^{-5}$, while the green one is for $|f_{\rm R0}|=10^{-6}$.
	The gray error bars represent the standard devotion of ensemble average of stacked signals over 100 realizations.
	\label{fig.haloestack}
	}
\end{figure}

We start to compare the result obtained from 100 simulations
for $\Lambda$CDM with the halo model as in Section~\ref{sec.stlen}.
The left panel in Figure~\ref{fig.haloestack} summarizes 
the stacked tangential shear profiles in each mass and redshift bin.
The black point in the left panel corresponds to the average values of tangential shear
over 100 realizations and the error bar represents the standard deviation of the average tangential shear 
(i.e. the standard deviation over 100 realizations divided by $\sqrt{100}$).
We also show the theoretical model as shown in Section~\ref{sec.stlen}
by the red lines. 
We have confirmed that our simulation results for the standard 
$\Lambda$CDM model are in good agreement with the halo model prediction for the range of $\theta = 1-100\, {\rm arcmin}$.

We then consider the comparison of tangential shear
between $f(R)$ gravity and $\Lambda$CDM.
In order to clarify the differences, we introduce 
the ratio of tangential shear between two cases.
The right panel in Figure~\ref{fig.haloestack} shows the ratio 
of tangential shear over 
three mass and redshift bins.
In the right panel, the red line corresponds to the ratio 
between the F5 model and $\Lambda$CDM,
while the green line is for the F6 model.
The gray error bars in the right panel represent the standard deviation 
of tangential shear for $|f_{\rm R0}|=0$.
According to the right panel in Figure~\ref{fig.haloestack},
there are found no differences between the stacked signal for 
the F6 and $\Lambda$CDM model 
over the wide range of halo masses and redshifts.
On the other hand, the differences from the $\Lambda$CDM model 
are clearly found in the case of the F5 model.
Interestingly, we find the significant deviation of the stacked signal
at $\theta \leq \theta_{\rm 200}$, where $\theta_{\rm 200}$ is the corresponding angular radius of $R_{200} = \left(3M_{\rm 200}/4\pi \, 200 \bar{\rho}_{m}\right)^{1/3}$.
This effect might be caused by the different mass assembly history
in $f(R)$ cosmology from the standard model, although 
the further investigation would be required.
Note that the recent numerical study
has reported that $f(R)$ gravity would induce the changes of the three-dimensional halo density profile \citep{2011PhRvD..83d4007Z, 2015arXiv151101494A}, 
while mass-concentration relation in F6 are found to be negligible for haloes with masses larger than $10^{13}\, h^{-1}M_\odot$ 
for all redshift \citep{2015MNRAS.452.3179S}.
We also find that the stacked signal at the transition scale 
between the one-halo and two halo terms would be affected by
the modification of gravity.
The trend of effects of $f(R)$ gravity on stacked profile
shows the complex dependences of the halo masses and redshifts.
We have confirmed that 
the three-dimensional halo density profile $\rho_{h}$ 
shows the similar trend as shown in Figure~\ref{fig.haloestack}
when we stacked $\rho_h$ for each halo mass and redshift bin,
i.e., more concentrated halo profile tends to make the tangential shear
larger at $\theta\simlt \theta_{200}$ and vice versa.

\subsubsection{Voids}
\label{sec.voidstack}

We next consider the stacked signals around voids.
In the stacked analysis, 
we divide the voids by their radius obtained from the {\tt Void Finder}. 
We consider two radius bins of 
$20-30\, h^{-1}{\rm Mpc}$ and $30-40\, h^{-1}{\rm Mpc}$.
Note that the total number of voids over 100 realizations is found 
to be $\sim1000$.
This means that we can find only $\sim10$ voids in each realization.
Since the stacked signals over $\sim10$ voids are quite weak,
it is difficult to discuss the impact of $f(R)$ gravity on the stacked signals
for each 25 ${\rm deg}^2$. 
Therefore, we decide to measure the stacked signal using all voids 
in 100 realizations.
The error bars of the stacked signals at the angular separation of $\theta$ are simply estimated by summing up the error which obtained from void profiles and LSS, and shape noise
\beqa
\sigma^2_{\gamma_{+}}=\sigma^2_{\rm void}+\sigma_{\rm LSS}^2+
\frac{\sigma_{e}^2}{2n_{\rm gal}}\left\{\frac{1}{N_{\rm stack}(2\pi\theta \Delta \theta)}\right\},
\label{eq:shape_noise_err}
\eeqa
where $N_{\rm stack}$ represents the number of objects used for stacked analysis and $\Delta\theta$ is the bin size of angular separation.
$\sigma_{\rm void}$ and $\sigma_{\rm LSS}$ are errors from dark matter distributions around voids and large-scale structure on the line of sight,
which are estimated from a covariance matrix with Eq.(11) in \citet{2013MNRAS.432.1021H}.
We here consider the 25 bins of angular separation 
in the range of 0-200 arcmin with 
$\Delta\theta = 8\, {\rm arcmin}$.

\begin{figure}
\centering
\includegraphics[width=0.45\columnwidth]{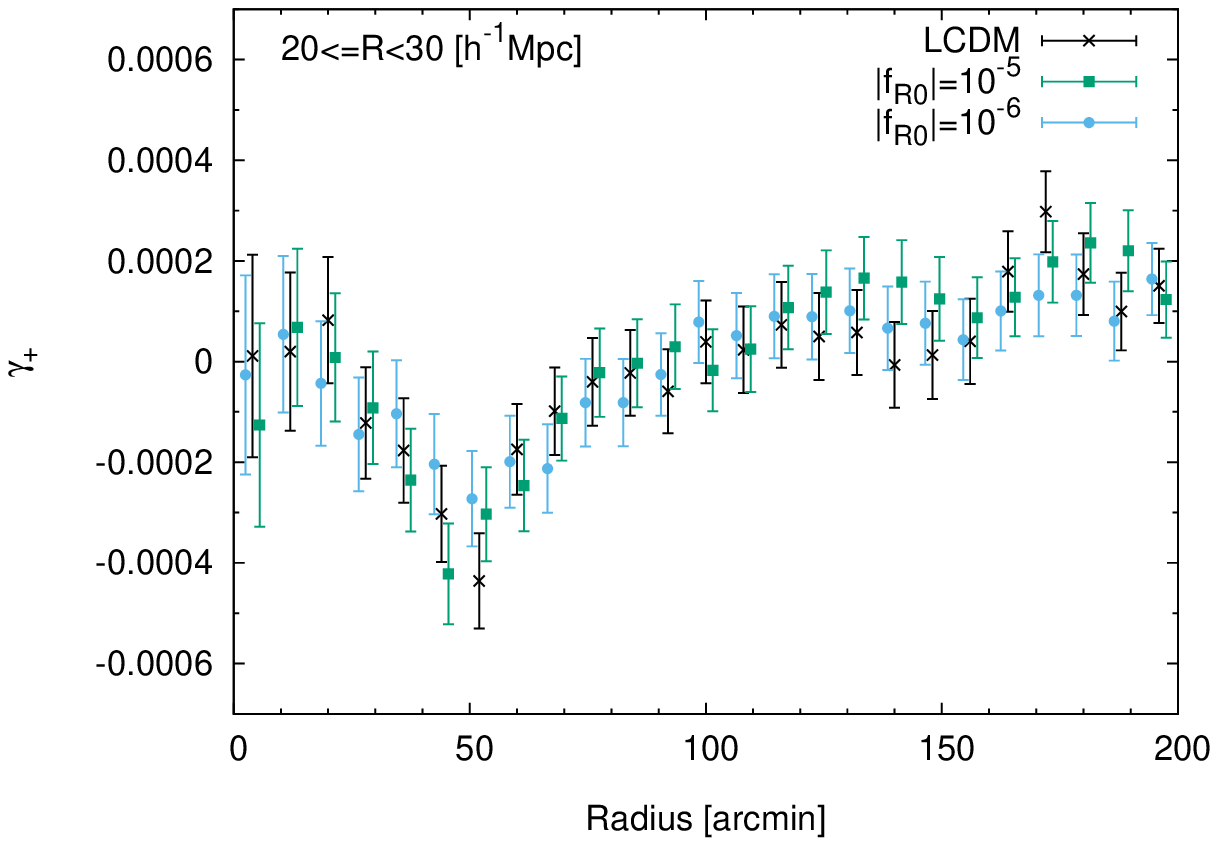}
\includegraphics[width=0.45\columnwidth]{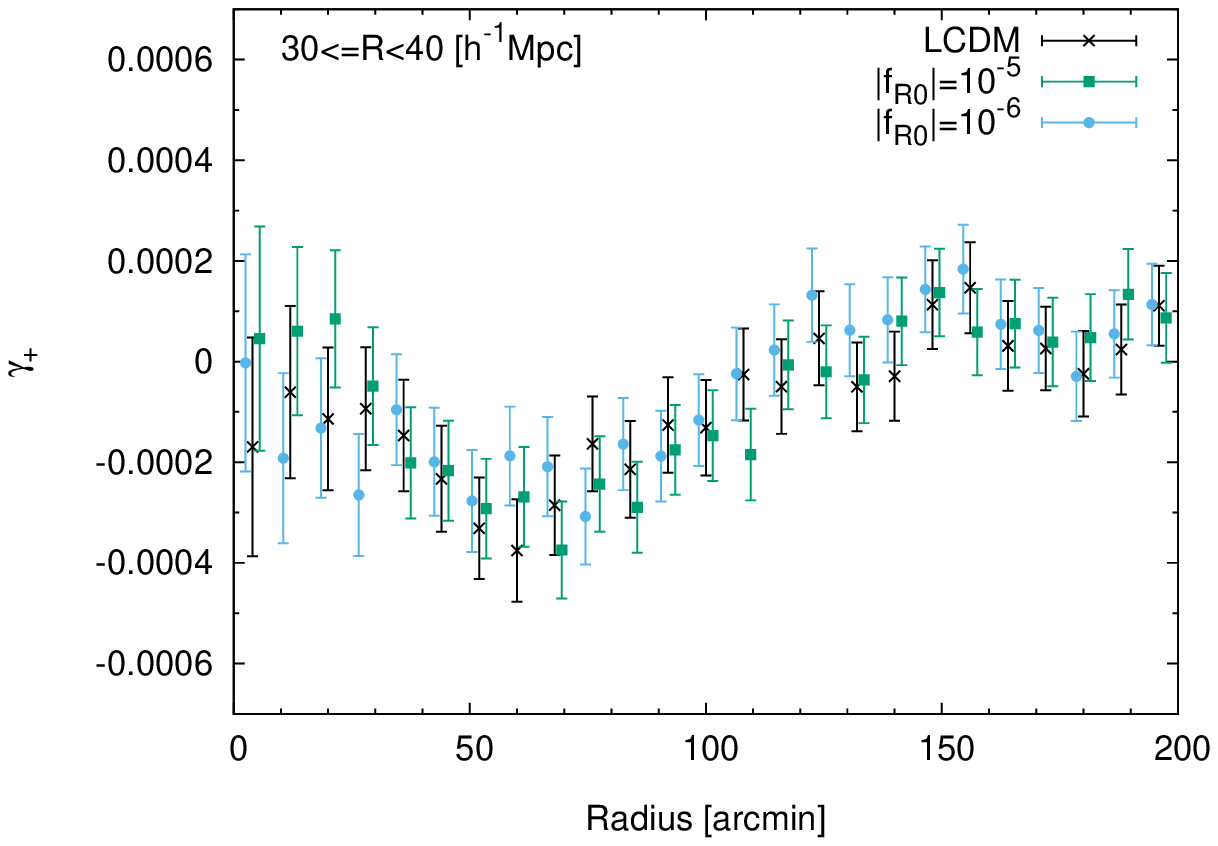}
\caption{
	Stacked tangential shear profiles around voids 
	for two bins of void radius $R$. 
	The horizontal axis shows angular distance from a center of a void.
	Crosses, squares and circles show the profiles for $\Lambda$CDM, 
	$|{\rm f_{R0}}|=10^{-5}$ and $|{\rm f_{R0}}|=10^{-6}$,
	respectively.
	The error bars represent the statistical 
	uncertainty related to the shape noise for a sky coverage of 
	2500 ${\rm deg}^2$.
	{\it Left}: $20\leq R\, [h^{-1}{\rm Mpc}] <30$ , 
	{\it Right}: $30\leq R\, [h^{-1}{\rm Mpc}]<40$
	\label{fig.stackvoidlin2}
	}
\end{figure}

Figure~\ref{fig.stackvoidlin2} shows the results of void stacking 
in each void radius bin as a function of distance from a center of a void. 
The error bar shown in this figure corresponds to 
the statistical uncertainty for a 2500 ${\rm deg}^2$ survey.
Clearly, it is difficult to find any effects on $f(R)$ gravity 
on the stacked signals around voids.
In order to evaluate the profile quantitively, 
we fit the profiles with void model as shown in Section~\ref{sec.stlen}.
Table~\ref{tab.fitvoid} shows the results of fitting with the double top-hat model. 
The $1\sigma$ uncertainty of fitting parameters in Table~\ref{tab.fitvoid}
are estimated with boot strap method.
We select voids randomly, stack and measure
the stacked profile over selected voids.
We then make $100$ stacked profiles for a given void radius bin
and calculate the standard deviation over 100 profiles. 
We perform the profile fitting by using Eq.~(\ref{eq:shear_void})
with three fitting parameters: density contrast of a void $\delta_1$, 
size of a void $\theta_1$ and size of a ridge $\theta_2$.  
In the fitting, lens redshift is set to be $z_l=0.4$.
We cannot find significant difference between estimated parameters, indicating that profiles are same within errors.

\citet{2015MNRAS.451.1036C} finds that stacking voids can constrain $f(R)$ gravity down to F5 with a Gpc scale survey.
In our analysis, on the other hand, it is difficult to find large differences between cosmological models for each size of voids.
This discrepancy mainly comes from the difference between stacking methods and the number of voids used in the stacking analysis.
They use more than $10^4$ voids for the stacking while we only use $10^3$ voids at most.
When their errors are scaled with the number of voids used in this paper, the errors are increased by more than three times.  
Therefore, one can not distinguish the profiles between $f(R)$ and $\Lambda$CDM when also using their simulation. 
This result is consistent with our result which includes more realistic effects by using ray-tracing simulations
As a result, stacked void lensing is a promising tool for giving constraint on $f(R)$ in future large-scale survey. 
However, larger surveys such as LSST \citep{2008arXiv0805.2366I} are required.

\begin{table}
\begin{center}
\caption{
Results of the fitting with the double top-hat model. 
The lens redshift is set to be 0.4. 
Errors show $1\sigma$ estimated with boot strap method.
Column (1): cosmological model, Column(2)-(3): fitting result. 
The unit of $\theta_1$ and $\theta_2$ is arcmin.}
\begin{tabular}{ccc}
\hline
model\ $\setminus$ radius [h$^{-1}$Mpc]&$20-30$&$30-40$\\ \hline\hline
&$\delta_1=-0.512\pm0.135$&$\delta_1=-0.472\pm0.119$\\
$\Lambda$CDM&$\theta_1=47.0\pm3.01$&$\theta_1=53.7\pm7.12$\\
&$\theta_2=81.6\pm29.5$&$\theta_2=122\pm34.7$\\ \hline

&$\delta_1=-0.688\pm0.153$&$\delta_1=-0.475\pm0.158$\\
$|{\rm f_{R0}}|=10^{-5}$&$\theta_1=40.9\pm3.46$&$\theta_1=57.9\pm10.6$\\
&$\theta_2=85.2\pm18.4$&$\theta_2=131\pm21.2$\\ \hline

&$\delta_1=-0.413\pm0.168$&$\delta_1=-0.363\pm0.231$\\
$|{\rm f_{R0}}|=10^{-6}$&$\theta_1=50.0\pm9.49$&$\theta_1=64.0\pm17.5$\\
&$\theta_2=90.7\pm13.8$&$\theta_2=112\pm22.0$\\ \hline
\hline
\end{tabular}
\label{tab.fitvoid}
\end{center}
\end{table}

\subsubsection{Troughs}
\label{sec.troughstack}

\begin{figure}
\begin{center}
\includegraphics[width=0.45\columnwidth]
{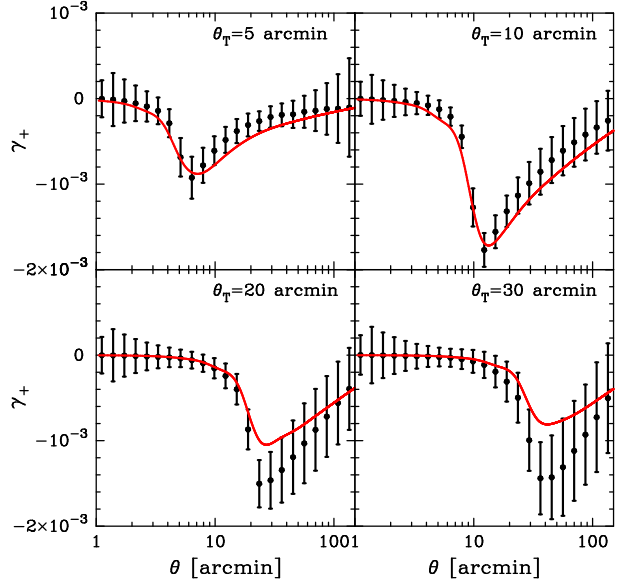}
\end{center}
\caption{
	The comparison between the measured stacked signals 
	around troughs from 100 $\Lambda$CDM simulations and 
	the theoretical prediction in Section~\ref{sec.stlen}.
	We consider the troughs selected from 
	the halo catalogs with 
	$M_{\rm T} = 10^{13}\, h^{-1}M_{\odot}$.
	The black point shows the average stacked signal over 
	100 realizations and the error bars show 
	the standard deviations estimated from 100 maps.
	The red line represents the theoretical model as proposed in
	\citet{2016MNRAS.455.3367G}.
}
\label{fig.comp_model}
\end{figure} 

\begin{figure}
\centering
\includegraphics[width=0.45\columnwidth]{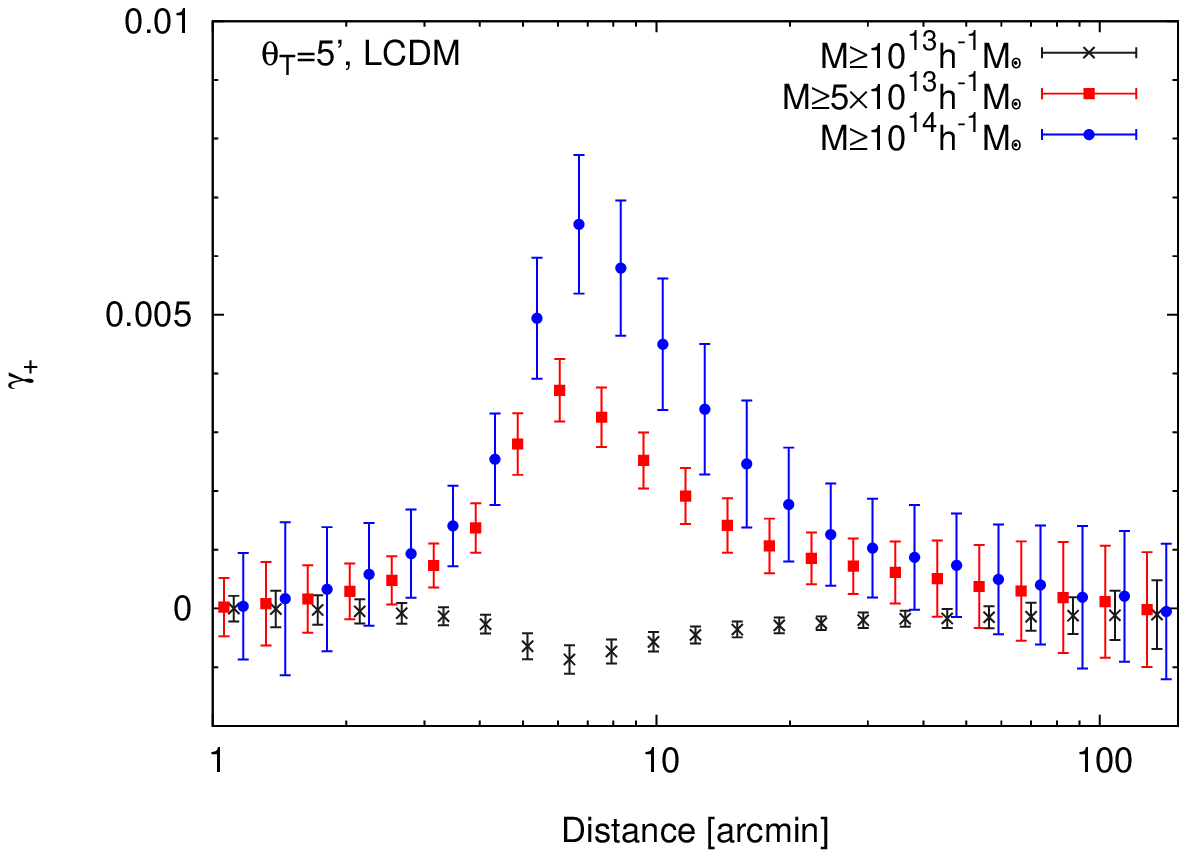}
\includegraphics[width=0.45\columnwidth]{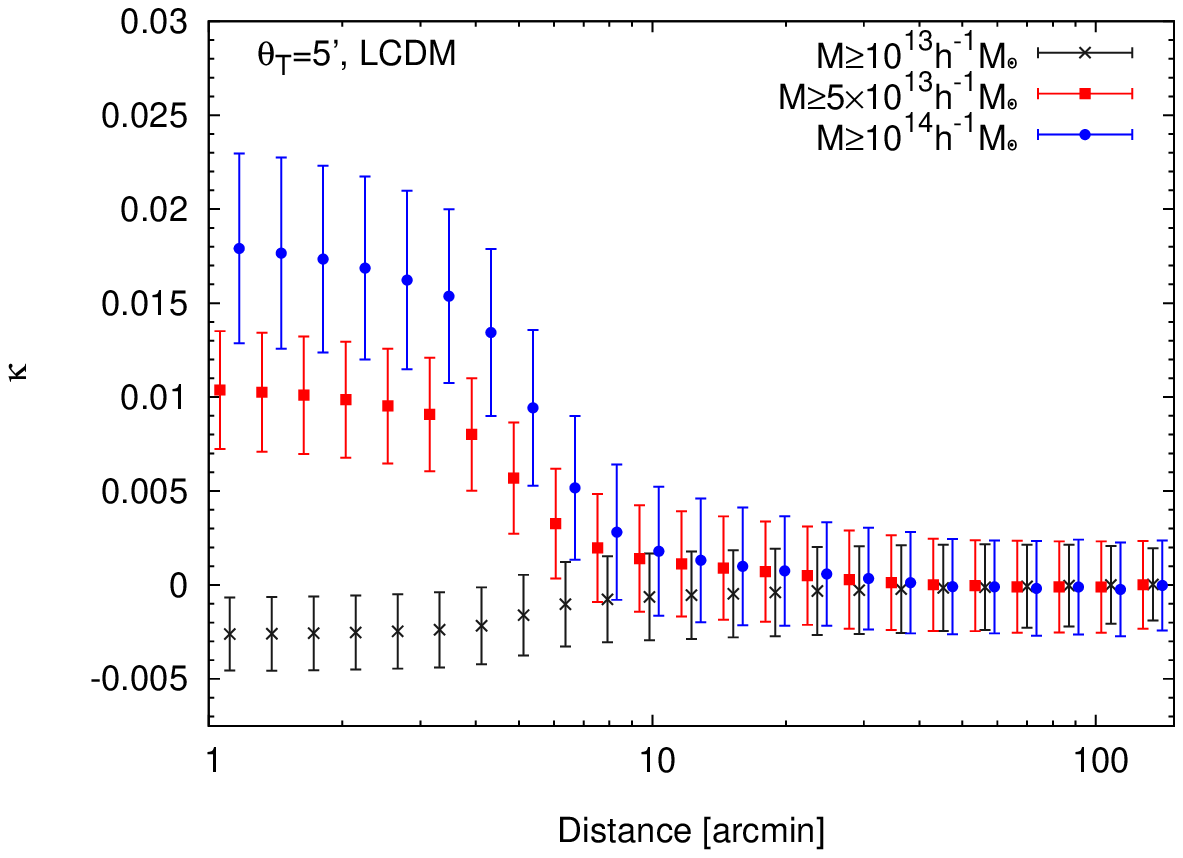}
\caption{
	Stacked signals around troughs with $\theta_{\rm T}=5'$ 
	in the $\Lambda$CDM model 
	when we apply the different halo mass criteria.
	Crosses, squares and circles show the profiles with mass criteria 
	$M_{\rm T}=10^{13}\, h^{-1}M_\odot$, 
	$5\times10^{13}\, h^{-1}M_\odot$ 
	and $10^{14}\, h^{-1}M_\odot$.
	{\it Left}: tangential shear profile, 
	{\it Right}: convergence profile.
	\label{fig.troughstack2}
}
\end{figure}

In this section, 
we evaluate the effects of $f(R)$ gravity 
on the stacked signal around troughs. 
We estimate the error bars of tangential shear
by the standard deviation over 100 realizations
and take into account the contribution from shape noise 
by using third term in Eq.~(\ref{eq:shape_noise_err}). 
Note that the errors represent the uncertainty with the sky coverage of $25$ square degrees.

We first compare the simulation results in the 
$\Lambda$CDM model with the theoretical model in \citet{2016MNRAS.455.3367G}.
Figure~\ref{fig.comp_model} summarizes the results in the case of the halo catalog with the mass threshold of 
$M_{\rm T}=10^{13}\, h^{-1}M_{\odot}$.
We consider the four cases of trough radius $\theta_{\rm T}=5, 10, 20$ and 30 arcmin.
The black point in Figure~\ref{fig.comp_model} represents 
the average stacked signals $\langle \gamma_{+} \rangle$
over 100 realizations, while the red line corresponds to the theoretical model.
We find that the model in \citet{2016MNRAS.455.3367G} provides 
the reasonable fit to the simulation results 
for $M_{\rm T}=10^{13}\, h^{-1}M_{\odot}$ 
and different four trough radiuses.
However, the measured signals around troughs would be strongly affected 
by the selection criteria of halo masses.
Figure~\ref{fig.troughstack2} shows the dependence of $M_{\rm T}$
on the stacked signals around the troughs in the case of $\Lambda$CDM.
The left panel in Figure~\ref{fig.troughstack2} summarizes 
the results of $\langle \gamma_{+}\rangle$ in the case of 
$\theta_{T}=5$ arcmin, while the right one corresponds to 
the average convergence profile $\langle \kappa \rangle$.
We find that the expected profiles as shown in \citet{2016MNRAS.455.3367G} are observed only 
when we set to be $M_{\rm T}=10^{13}\, h^{-1}M_{\odot}$ 
in our simulations. 
Convergence profiles with large halo mass criteria tend to show positive values 
while this trend appears for lower mass selection criteria.
This might be caused by 
the presence of massive halos with mass of $M<M_{\rm T}$
in a trough radius.
However, this effect is found to be smaller as the trough radius increases.

\begin{figure}
\centering
\includegraphics[width=0.45\columnwidth]{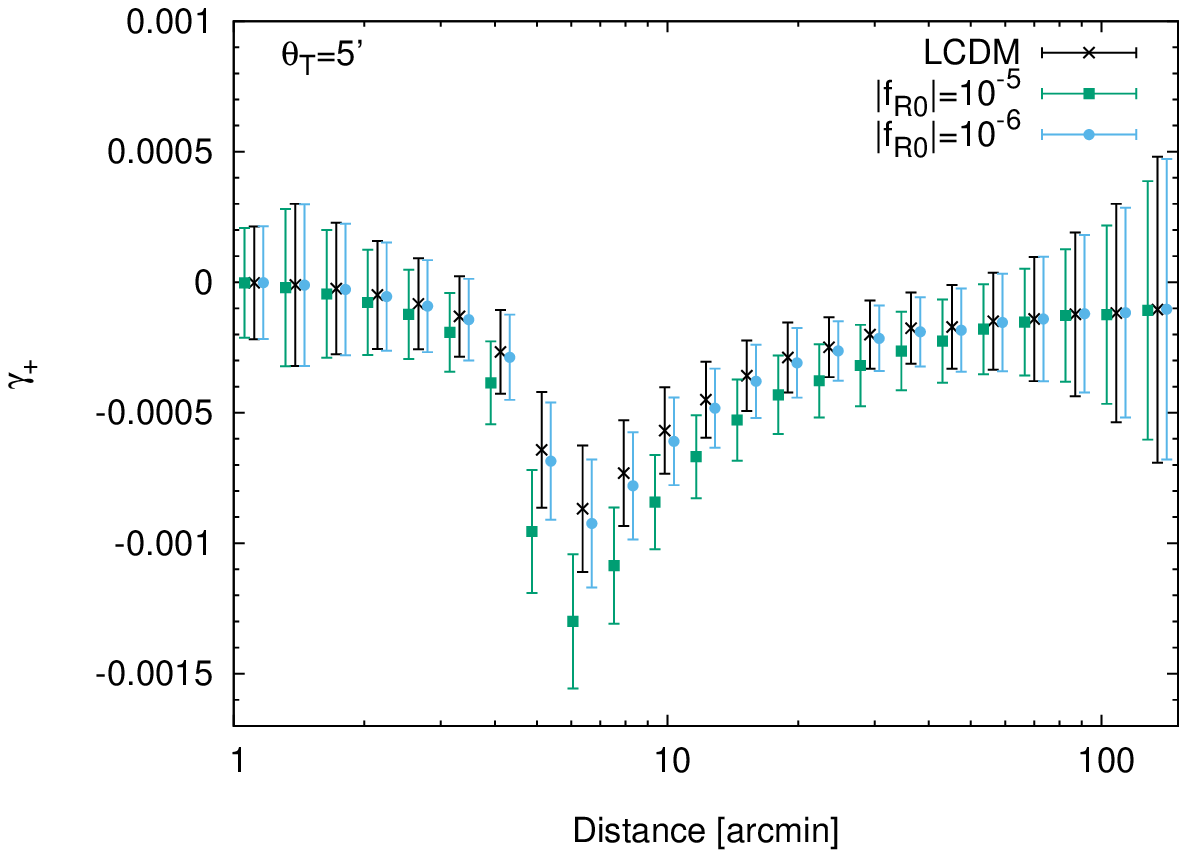}
\includegraphics[width=0.45\columnwidth]{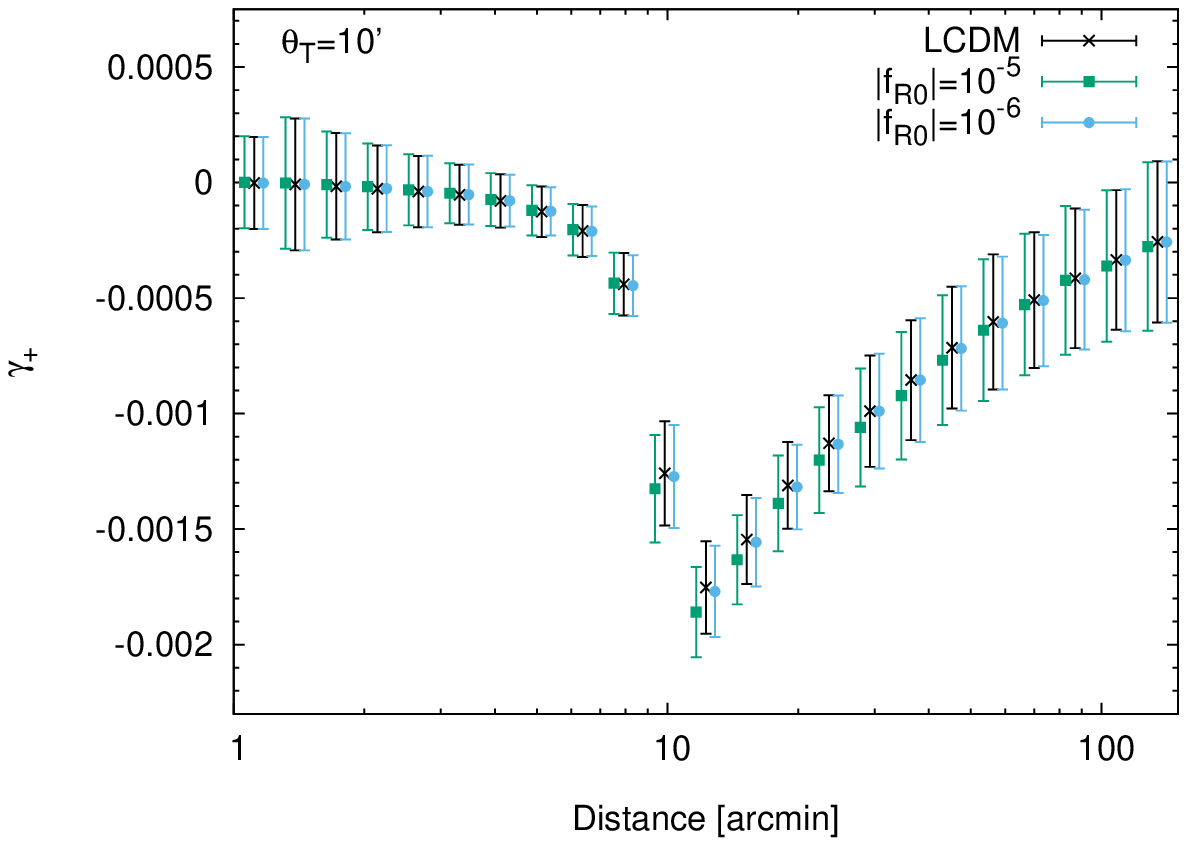}
\caption{
	Stacked profiles of troughs for two bins of trough radius. 
	We set halo masses to be 
	$M_{\rm T}=10^{13}\, h^{-1}M_{\odot}$ 
	for the selection of troughs. 
	Crosses, squares and circles show the profiles for $\Lambda$CDM, 
	$|{\rm f_{R0}}|=10^{-5}$ and $|{\rm f_{R0}}|=10^{-6}$.
	{\it Left}: $\theta_{\rm T}=5$ arcmin, 
	{\it Right}: $\theta_{\rm T}=10$ arcmin}
	\label{fig.troughstack}
\end{figure}

We then consider the imprint of $f(R)$ gravity on the stacked signals
around troughs.
Figure~\ref{fig.troughstack} shows the stacked tangential shear profiles for the $\Lambda$CDM and two $f(R)$
models. 
As seen in this figure, we can find the differences of 
$\langle \gamma_{+} \rangle$ at $\theta\sim\theta_{\rm T}$
between the F5 model and $\Lambda$CDM model
when using the troughs with $\theta_{T}=5$ arcmin and 
the halo catalogs with $M_{\rm T}=10^{13}\, h^{-1}M_{\odot}$.
In order to define the significance with the more quantitive manner,
we introduce the significance level of the difference 
between the two models of tangential shear profile 
$\gamma_{+,(a)}$ and $\gamma_{+,(b)}$ for each bin as follows:
\beqa
(S/N)^2 =  
\sum_{i} 
\frac{\left[\gamma_{+,(a)}(\theta_{i})-\gamma_{+,(b)}(\theta_{i})\right]^2}{\sigma^2 N_{bin}^2},
\label{eq:SNR_profile}
\eeqa
where $\sigma$ represents the statistical uncertainty in stacked analysis
for a given sky coverage.
In order to calculate Eq.~(\ref{eq:SNR_profile})
for the troughs with $\theta_{T}=5$ arcmin and 
$M_{\rm T}=10^{13}\, h^{-1}M_{\odot}$,
we consider 23 bins in the range of 1-150 arcmin and
compute the standard deviation of stacked signals over 100 
$\Lambda$CDM simulations.
We then use this standard deviation as the estimator of $\sigma$ 
in a 25 ${\rm deg}^2$ sky.
Assumed that $\sigma$ would be scaled with survey area,
we find that the signal-to-noise ratio 
between the F5 model and $\Lambda$CDM model
would be 1.43 for each bin in a sky coverage 
of 1,400 ${\rm deg}^2$ as proposed in the ongoing Subaru/Hyper Suprime-Cam (HSC) survey \citep{2006SPIE.6269E...9M}.
Therefore, we conclude that the stacked signal around troughs
can be useful to distinguish the $f(R)$ model with 
$|f_{\rm R0}|=10^{-5}$ and 
the standard $\Lambda$CDM model
with a $\sim2\sigma$ level in the ongoing large-scale survey.
Note that we cannot observe any significant differences of 
tangential shear between the F5 model 
and $\Lambda$CDM model
over 100 realizations for a set of $M_{\rm T}$ and $\theta_{\rm T}$,
except for the case of 
$\theta_{T}=5$ arcmin and
$M_{\rm T}=10^{13}\, h^{-1}M_{\odot}$.
Moreover, we cannot find any significant differences between 
the stacked profile around troughs in the F6 model and the $\Lambda$CDM model in our simulations.

\subsection{Peak Statisitcs}
\label{sec.peakstat}

In this section, 
we simulate galaxy shape noise in our simulation by
adding the shape noise to shear from random ellipticities
which follow the two-dimensional Gaussian distribution 
as 
\beqa
P(|e|)=\frac{1}{\pi\sigma_{\rm int}}
\exp\left(-\frac{e^2}{\sigma_{\rm int}^2}\right),
\eeqa
where $\sigma^2_{\rm int} = \sigma_{\rm e}^2/(n_{\rm gal}\theta^2_{\rm pix})$ with 
the pixel size of $\theta_{\rm pix}=0.075$ arcmin.
When analyzing the peaks, we remove the regions 
within smoothing scale $\theta_{\rm G}$ 
from edges of each convergence map.

\subsubsection{Peak-Halo matching}

\begin{figure}
\centering
\includegraphics[width=0.43\columnwidth]
{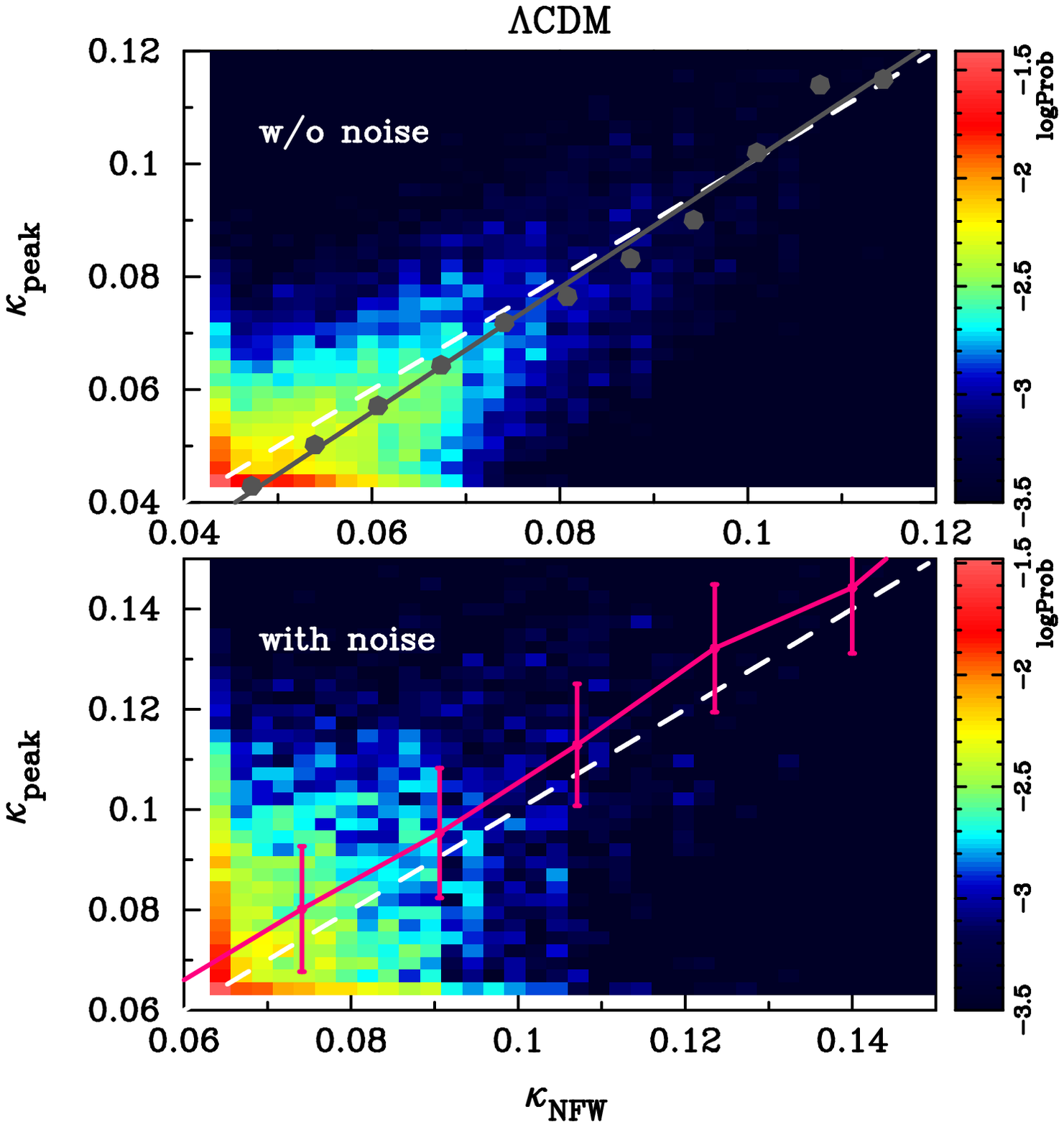}
\includegraphics[width=0.43\columnwidth]
{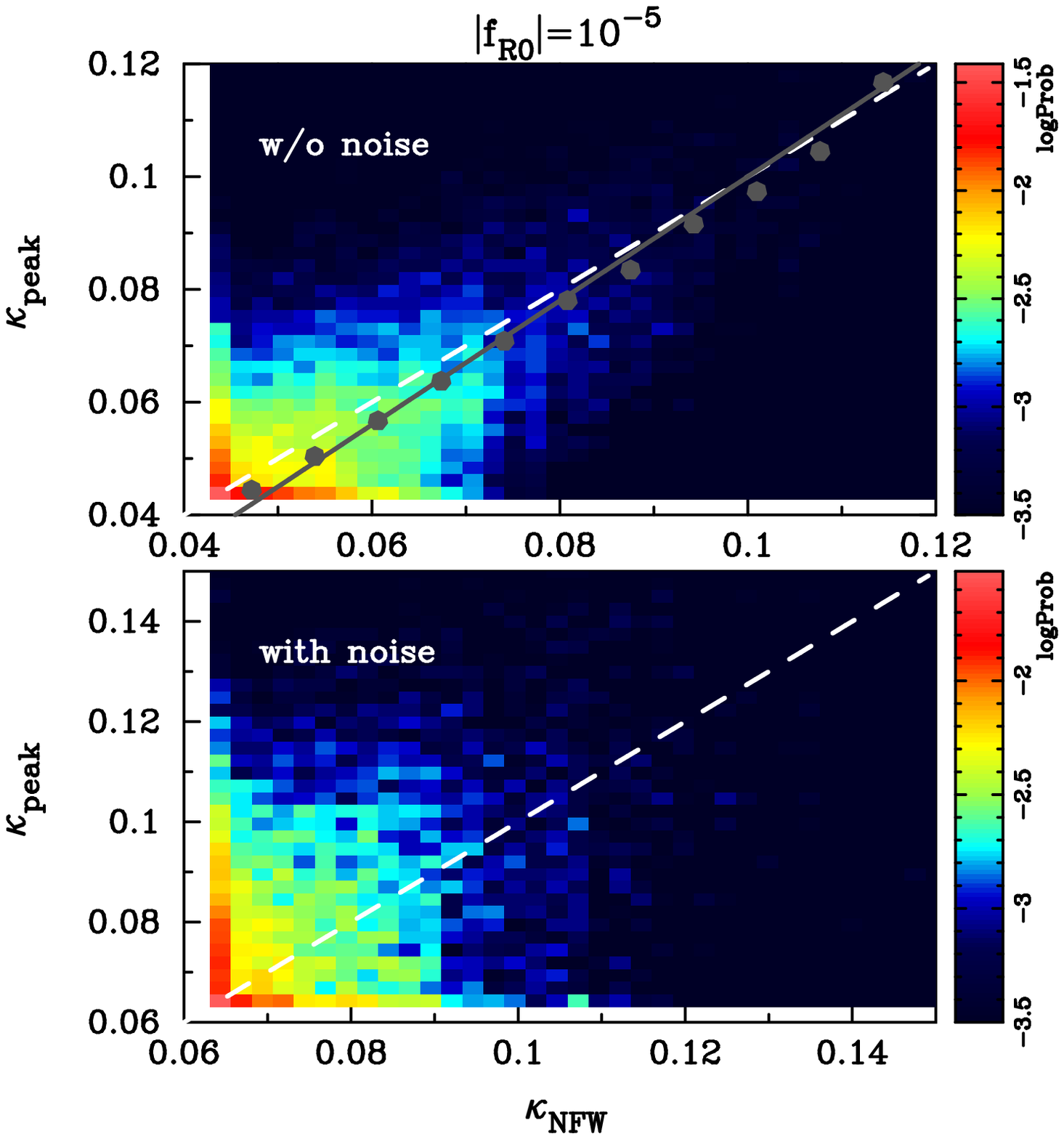}
\caption{
	The correspondence between dark matter haloes and lensing peaks.
	In each panel, the vertical axis represents the peak height and 
	the horizontal axis shows the expected convergence of NFW haloes.
	In both right and left panels, 
	the top panel shows 
	the scatter plot of peak height and the expected convergence by
	the matched haloes in absence of noise, while
	the lower panel corresponds to the case with shape noise.
	{\it Left}: The $\Lambda$CDM model.
	{\it Right}: $f(R)$ model with $|f_{\rm R0}|=10^{-5}$.
	\label{fig:halo_peak_comp}
}
\end{figure}

We first examine the correspondence between dark matter haloes 
and the local maximum in smoothed convergence map.
With our ray-tracing simulations and mock halo catalogs, 
we study the correspondence between haloes and the peaks 
in weak lensing convergence maps.
We first identify the local maxima 
in the smoothed convergence field 
with source redshift of $z_{\rm source}=1$.
For selection of peaks, 
the threshold of peak height is set to be 
${\cal K}=0.04$ for noise-less maps
and 0.06 for noisy maps.
These value correspond to $\sim3\sigma$ 
in smoothed convergence maps without and with noise, respectively.
For a given position of lensing peak, 
we search for the matched dark matter haloes 
within a radius of 3 arcmin from the peak position.
This search radius is set to be larger than the smoothing scale 
but still smaller than the angular size of massive haloes at $z\sim 0.1-0.7$
\citep[also see,][]{2004MNRAS.350..893H}.
When we find several haloes in search radius, 
we regard the matched halo as 
the closest halo from the position of peak.
For each matched peak, we estimate the corresponding convergence by using the universal NFW density profile (Eq.~\ref{eq:NFW}).
In the calculation of expected convergence from haloes, 
we convert the mass defined by $M = 4\pi/3 \times 200\bar{\rho}_{m}R_{\rm 200}^3$ to the virial mass
by using the halo concentration shown in Eq.~(\ref{eq:cvir_model}).
In total, for the $\Lambda$CDM cosmology,
we find 25,806 and 12,865 pairs of peaks and haloes 
over 100 noise-less maps and noisy maps.

The left panels of Figure~\ref{fig:halo_peak_comp} 
shows the scatter plot of peak height in convergence map
and the expected convergence by NFW haloes in the case of
the $\Lambda$CDM cosmology.
The vertical axis corresponds to peak height, 
while the horizontal axis shows the corresponding convergence 
expected by NFW haloes.
Thus, the colour map in each panel 
shows the probability of 
${\rm Prob}({\cal K}_{\rm peak, obs}|{\cal K}_{{\rm peak}, h})$.
We present the line of $y=x$ as the dashed line in each panel.
In lower panel, 
we show the effect of the modulation of peak height 
as the magenta line with error bars.
The magenta line represents 
$\langle {\cal K}_{\rm peak, obs} | {\cal K}_{{\rm peak}, h} \rangle$,
which is defined by 
\beqa
\langle {\cal K}_{\rm peak, obs} | {\cal K}_{{\rm peak}, h} \rangle
(z, M) 
= \int {\rm d}{\cal K}\, {\cal K}\,  
{\rm Prob}({\cal K}|{\cal K}_{{\rm peak}, h}(z, M))
\label{eq:mean_peak_height}
\eeqa
and the error bars reflect the scatter of
$\langle {\cal K}_{\rm peak, obs} | {\cal K}_{{\rm peak}, h} \rangle$.
As shown in previous works, 
we confirm the good correspondence between the matched dark matter haloes and lensing peaks in the noise-less maps.
Nevertheless, even in the noise-less maps, 
the better correspondence between the matched haloes and 
peaks would be generalized by \citep{2015MNRAS.453.3043S}
\beqa
{\cal K}_{\rm peak, obs}(z, M) = 
c_{0} + c_{1} {\cal K}_{{\rm peak}, h}(z, M).
\label{eq:mean_peak_height_wonoise}
\eeqa
In the top left panel of Figure~\ref{fig:halo_peak_comp},
the gray line shows Eq.~(\ref{eq:mean_peak_height_wonoise}) 
with $c_{0}=-0.01$ and $c_{1}=1.1$ and 
the gray point represents $\langle {\cal K}_{\rm peak, obs} | {\cal K}_{{\rm peak}, h} \rangle$ measured from the simulation results.
Also, our model as shown in Eq.~(\ref{eq:mean_peak_height}) 
can explain the average relation 
between peaks and dark matter haloes even in the case with noise.

We also perform the similar calculations in the F5 model.
We simply assume the NFW profile with
the halo concentration expressed by Eq.~(\ref{eq:cvir_model})
in the case of non-zero $|f_{\rm R0}|$.
In the F5 model,
we find 29,767 and 14,410 pairs of peaks and haloes 
over 100 noise-less maps and noisy maps, respectively.
The right panels in Figure~\ref{fig:halo_peak_comp}
correspond to the case of the F5 model.
Compared to the left panels, 
we cannot find any significant impact on $f(R)$ gravity
on the correspondence between haloes and peaks in both noise-less maps and noisy maps.
In the right top panel, we again show
measured $\langle {\cal K}_{\rm peak, obs} | {\cal K}_{{\rm peak}, h} \rangle$ from the pairs of haloes and peaks by the gray points, 
while the gray line represents
Eq.~(\ref{eq:mean_peak_height_wonoise}) with 
$c_{0}=-0.01$ and $c_{1}=1.1$.
Thus, the probability of ${\rm Prob}({\cal K}_{\rm peak, obs}|{\cal K}_{{\rm peak}, h})$ is less affected by the modification of gravity
for the HS model with $|f_{\rm R0}|\simlt10^{-5}$.
This result is consistent with the result shown in Figure~\ref{fig.haloestack}
because the expected peak height for a given NFW halo 
${\cal K}_{{\rm peak}, h}$ would be determined mainly 
by the tangential shear profile at $\theta\sim2-3\theta_G$,
where the effect of $f(R)$ gravity should be less than 10\%.

\subsubsection{Abundance of high significance local maxima}

\begin{figure}
\centering
\includegraphics[width=0.45\columnwidth]
{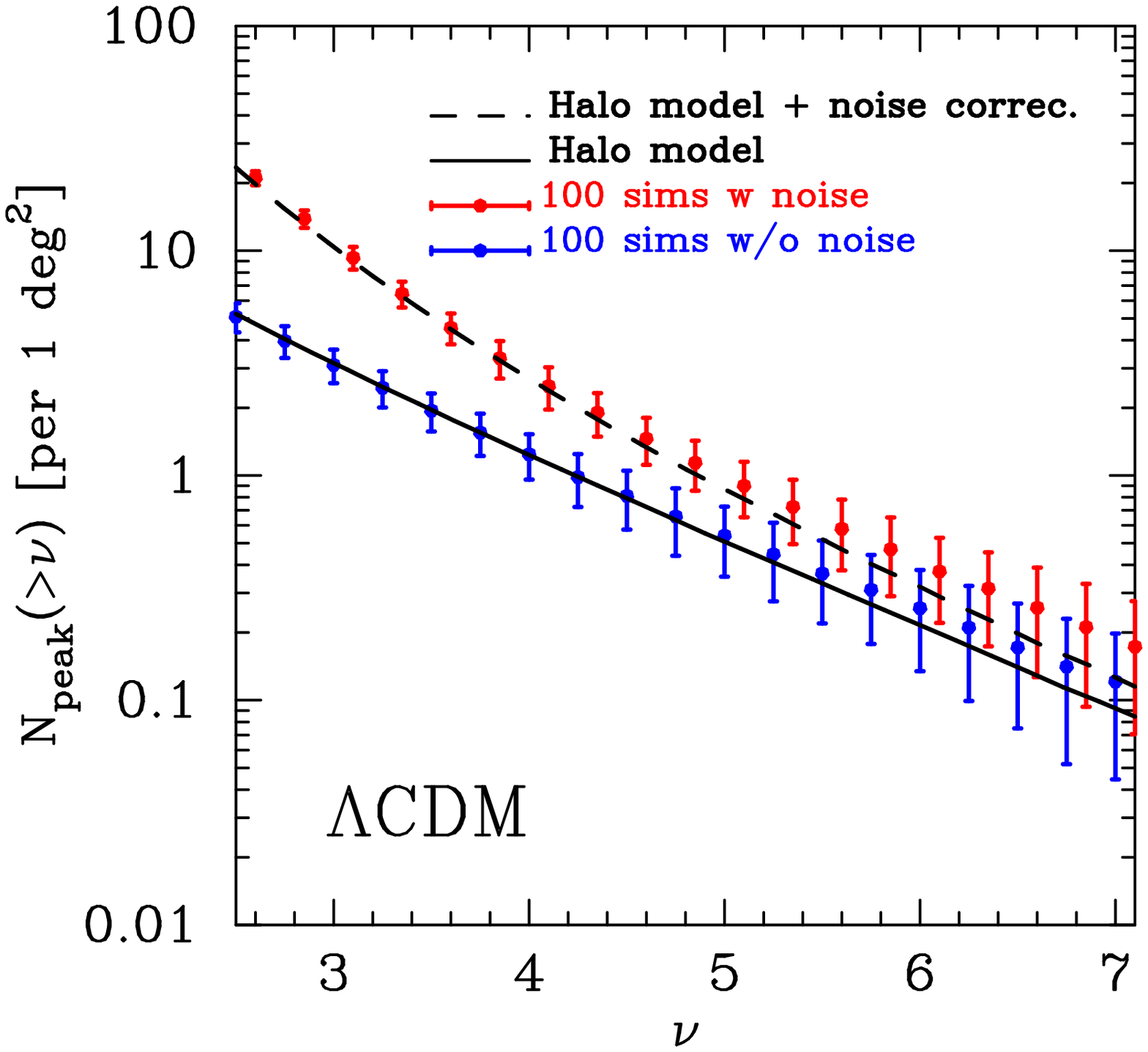}
\includegraphics[width=0.45\columnwidth]
{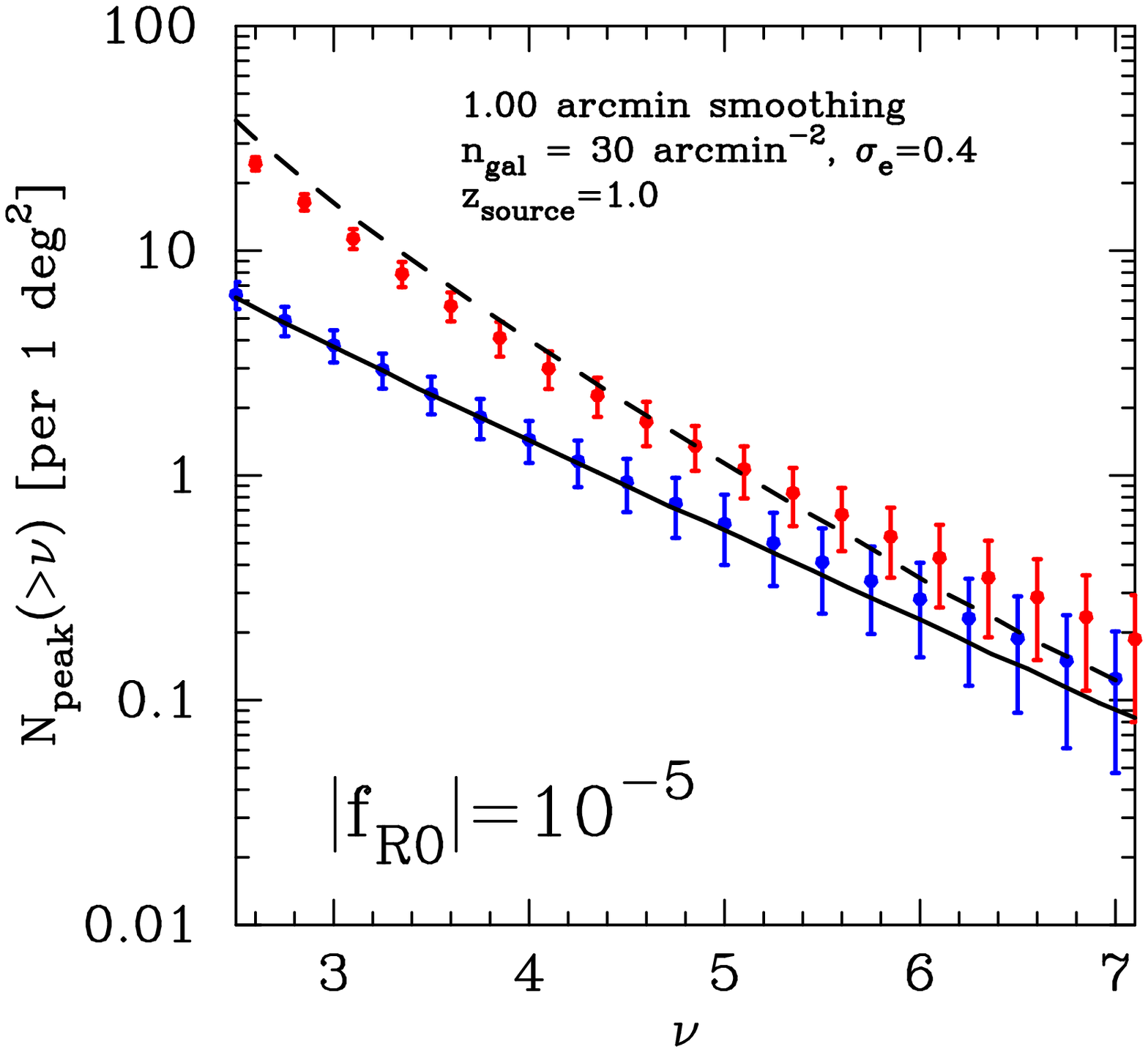}
\caption{
	The comparison of abundance of local maxima measured 
	in simulations and the halo model prediction.
	In both panels, the red points with error bars represent
	the results in 100 noisy maps, while the blue is for 
	the noise-less maps.
	The solid and dashed lines are our model with and without shape 	noises, respectively.
	{\it Left}: The $\Lambda$CDM model.
	{\it Right}: $f(R)$ model with $|f_{\rm R0}|=10^{-5}$.
	\label{eq:npeak_high}
}
\end{figure}

We then consider the abundance of local maxima with high height,
which is expected to be associated with single 
massive dark matter halo along a line of sight.
As shown in Figure~\ref{fig:halo_peak_comp},
the probability distribution function of observed peak height 
for a given matched dark matter halo 
is less affected by the modification of gravity.
This implies that the number count of high significance local maxima
can be useful to extract information about 
the number density of dark matter haloes, 
or the halo mass function.

Assumed that the halo density profile would not be 
affected by the modification of gravity, 
we can predict the abundance of high significance local maxima
as shown in Section~\ref{subsec:peak}.
Figure~\ref{eq:npeak_high} shows the comparison between measured abundance from 100 ray-tracing simulations
and our model.
In both left and right panels, the red points with error bars 
represent the number count of local maxima in the noisy maps,
while the blue is for the noise-less maps.
The error bars correspond to the standard deviation over 100 realizations.
Also, the dashed and solid lines are our model for the noisy and noise-less cases, respectively.

As shown in this figure, 
our model can provide the good fit to simulation results 
in the standard $\Lambda$CDM models, 
regardless of the presence or absence of shape noise.
For the $f(R)$ model, 
we compute the mass function with the {\it corrected} mass variance,
which is given by \citep{2011PhRvD..84h4033L}
\beqa
\sigma(M) = \frac{\sigma_{f(R)}(M) + \left(M/M_{\rm th}\right)^{p}\sigma_{\Lambda CDM}}{1+\left(M/M_{\rm th}\right)^{p}},
\label{eq:mass_variance_fR}
\eeqa
where $\sigma_{f(R)}$ is the mass variance with 
the linear matter power spectrum in the $f(R)$ model 
and $\sigma_{\Lambda CDM}$ is the mass variance in the $\Lambda$CDM model.
The transition mass scale $M_{\rm th}$ and 
the parameter $p$ are found to be
$M_{\rm th} = 1.345\times 10^{13}\, h^{-1}M_{\odot}
\left(|f_{\rm R0}|/10^{-6}\right)^{3/2}$ and $p=2.448$ 
in \citet{2011PhRvD..84h4033L}.
We take Eq.~(\ref{eq:mass_variance_fR}) and 
simply adopt the functional form of mass function calibrated 
in \citet{Tinker:2008ff}.
Note that we have confirmed that this is a reasonable approximation
in our halo catalogs with $M_{200}\ge10^{13}\, h^{-1}M_{\odot}$
and the redshift of 0-1.
Under these assumption, we can compare the abundance of high significance peaks with our model prediction and then find that 
the our model works even in the F5 model 
for both noisy and noise-less maps.

\subsubsection{General peak count and its cosmological application}

As summarized in Section~\ref{subsec:peak}, 
peaks can be defined by local maxima or minima in 
the smoothed convergence map in general.
In this section, we examine the effect on $f(R)$ gravity 
on number count of local maxima and minima by
using 100 ray-tracing simulations.
We estimate the statistical uncertainty of peak counts 
in a 25 ${\rm deg}^2$ area 
from the standard deviation over 100 realizations.
For the statistical uncertainty in the ongoing HSC survey
with the sky coverage of 1400 ${\rm deg}^2$, 
we scale the uncertainty with the sky coverage 
(i.e., by a factor of $\sqrt{1400/25}$).
\begin{figure}
\includegraphics[width=0.5\columnwidth]{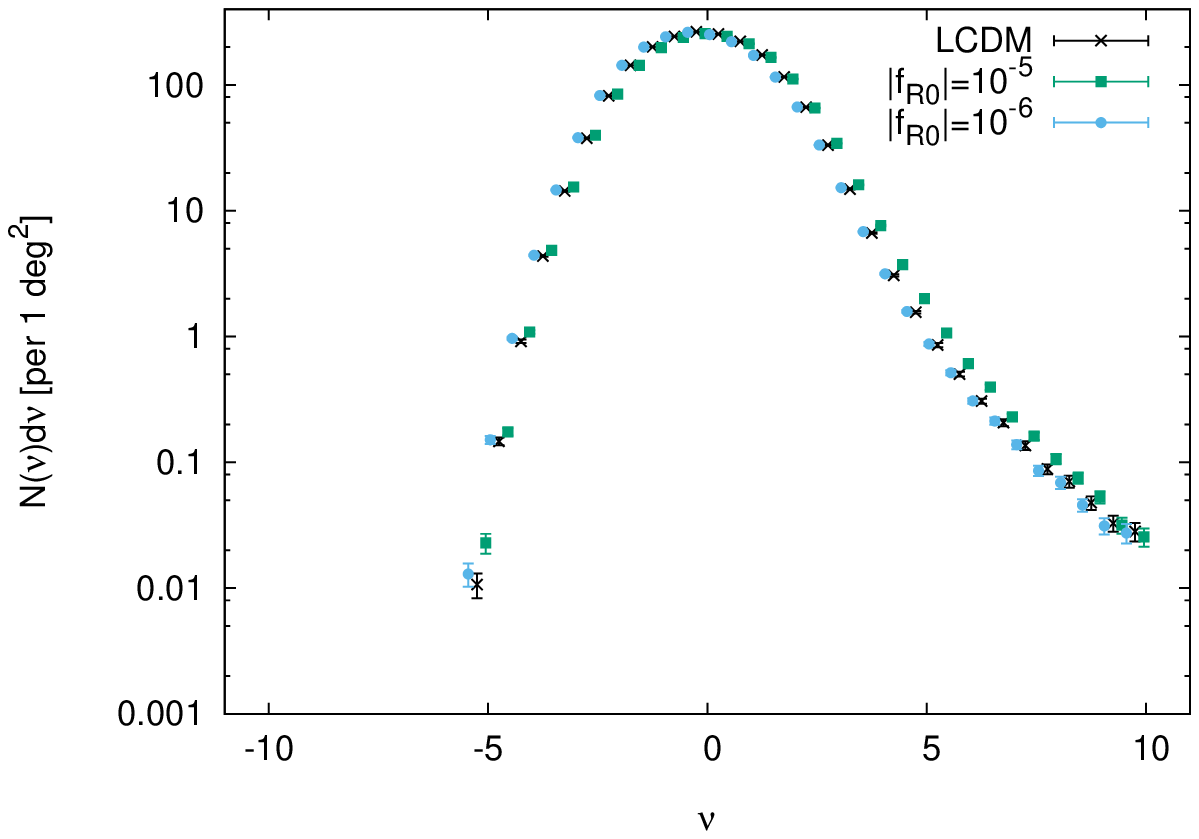}
\includegraphics[width=0.5\columnwidth]{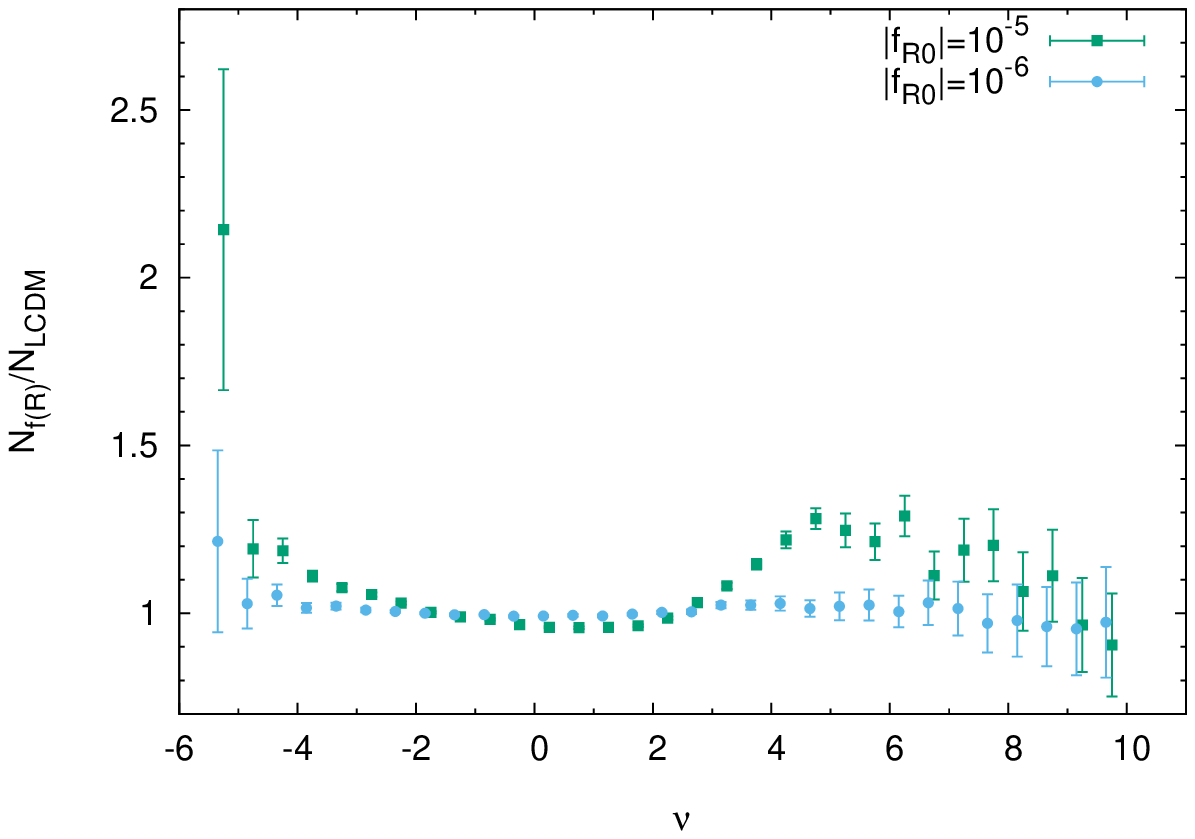}
\caption{
	Comparison of the number count of peaks 
	for each cosmological model. 
	The horizontal axis shows the peak height 
	normalized by the shape noise variance. 
	In both panels, we consider the number count of
	local maxima and minima in a convergence map
	with the bin size of $\Delta\nu=0.5$.
	In the left panel, crosses, squares and circles 
	show the number of lensing peaks for 
	$\Lambda$CDM, 
	$|{\rm f_{R0}}|=10^{-5}$ and 
	$|{\rm f_{R0}}|=10^{-6}$, respectively. 
	The right panel represents 
	the ratio of the number counts for the $f(R)$ model 
	normalized by that in the $\Lambda$CDM.
        The error bars correspond to
	the statistical uncertainty  
	for the ongoing HSC survey with 
	the sky coverage of 1400 ${\rm deg}^2$. 
	\label{fig.pdf1}
	}
\end{figure}

The left panel in Figure~\ref{fig.pdf1} shows the number count of peaks 
with the smoothing scale $\theta_{\rm G}=1$ arcmin and 
$\theta_{o}=10$ arcmin, while the right corresponds to
the ratio of the number of peaks between 
the $f(R)$ model and the $\Lambda$CDM model. 
The peak height $\nu$ are binned into $40$ bins 
from $-10$ to $10$ linearly.
For the HS model with $|f_{\rm R0}|=10^{-5}$,
we find that the number of peaks with the $\nu\simgt3$ would increase
with a level of $\sim30\%$ compared to the $\Lambda$CDM model.
Interestingly, the number of peaks with $\nu\simlt-2$ would
be also affected by the presence of the fifth force, 
increasing the number of peaks by a factor of $\sim1.5$.
Since negative peaks are caused by superposition of underdense regions on line of sight,
increasing the number of voids in $f(R)$ gravity might affect the number count of negative peaks.
On the other hand, we cannot find any significant deviations 
in peak counts for $|f_{\rm R0}|=10^{-6}$.

Let us quantify the significance of the differences shown in 
Figure~\ref{fig.pdf1} by introducing the following statistic
\beqa
\left(\frac{\rm S}{\rm N}\right)_{{\rm peak}}^2=
\sum_i
\left(\frac{N_{{\rm peak}, i}^{|f_{\rm R0}|}
-N_{{\rm peak}, i}^{\Lambda {\rm CDM}}}{\sigma_{{\rm peak}, i}}\right)^2,
\eeqa
where $N_{{\rm peak}, i}^{|f_{\rm R0}|}$ represents 
the number count of $i$-th bin for the model 
with the parameter of $|f_{\rm R0}|$,
$\sigma_{{\rm peak}, i}$ is the statistical uncertainty of the number count of $i$-th bin for $\Lambda$CDM.
Note that $N_{{\rm peak}, i}^{\Lambda {\rm CDM}}$ corresponds to the case of $\Lambda$CDM.
Table~\ref{tab.pdfsnscaled} shows the result of signal-to-noise ratio.
In the estimation, we do not use high peaks whose number count is zero.
The significance for the F5 model becomes more than $1\sigma$ as average value in each bin.

\begin{table}
\begin{center}
\caption{
	Total S/N which indicates difference of the number count of 
	peaks between $\Lambda$CDM and $f(R)$. 
	Error bars are estimated with 100 realizations and 
	scaled for the HSC survey. 
	Column (1): cosmological model, Column(2): 
	values of peaks $\nu$ used for the estimation
	\label{tab.pdfsnscaled}
}
\begin{tabular}{ccc}
\hline
$f(R)$ model & $-10\leq\nu\leq10$ (31 bins) & $|\nu|\ge4$ (15 bins) \\ \hline\hline
$|{\rm f_{R0}}|=10^{-5}$&$39.39$ & $20.81$ \\ 
$|{\rm f_{R0}}|=10^{-6}$&$7.269$ & $2.770$ \\ 
\hline
\end{tabular}
\end{center}
\end{table}

\section{CONCLUSION AND DISCUSSION}
\label{sec.con}

We have studied the matter density distribution 
in $f(R)$ gravity.
In particular, we have focused 
on the large-scale structures which can be observed in weak lensing measurement.
We have performed ray-tracing simulations in two different 
$f(R)$ gravity models and the standard $\Lambda$CDM model.
Using a set of weak lensing maps and dark matter halo catalogs,
we have investigated 
the connection with observables and large-scale
structures in a realistic way.
Throughout this paper, 
we have considered two different statistical methods 
and compared the statistics in the standard $\Lambda$CDM model
with the $f(R)$ model proposed in \citet{2007PhRvD..76f4004H}.
Our main findings are summarized as follows:

\begin{enumerate}

\vspace{2mm}
\item 
The averaged tangential shear profile around dark matter haloes
shows a clear dependence on $f(R)$ gravity.
The significant deviation from $\Lambda$CDM is found 
at {\it both} the inner region and the outskirt of dark matter haloes.
The trend of deviation would depend on the halo masses and redshifts
in a non-monotonic way.

\vspace{2mm}
\item 
While the averaged tangential shear profile around voids 
is expected to be a promising target to probe the modification of gravity,
we cannot find any significant differences between the $f(R)$ model
and $\Lambda$CDM model even if using $\sim1000$ voids.
This result indicates that the uncertainty of the center of voids 
and the projected matter distribution along a line of sight would mitigate the effect of $f(R)$ gravity on the signal around voids.
However, larger surveys would be able to distinguish profiles between $\Lambda$CDM and $f(R)$ gravity.

\vspace{2mm}
\item 
Troughs, the underdense regions in the surface density field 
of galaxies, are proposed as the tracer of the underdensity in the Universe.
In this paper, we consider the troughs defined by dark matter haloes
and examine how the tangential shear profile around troughs 
would be affected by the modification of gravity.
The stacked tangential shear profile around troughs 
clearly shows the contribution from the underdensity in the Universe. 
The clear deviation from $\Lambda$CDM have been confirmed 
when the angular separation would be equal to 
the search radius of troughs.
However, we also find that correct understanding 
of halo catalogs would be required in order to 
sample the underdense regions in the Universe with troughs.

\vspace{2mm}
\item
Peaks in a reconstructed mass map from weak lensing measurement
can in principle contain the cosmological information 
about dark matter haloes, voids and structures along a line of sight.
We perform matching analysis of local maximum in a map 
to dark matter halo along the line of sight.
We then confirm that the clear correspondence between local maxima 
and haloes in absence of shape noises and 
the modulation effect in the relation of maxima and haloes 
for noisy maps.
Therefore, proper understanding of the relation of maxima and haloes
enables us to extract the information about the halo mass function.
We demonstrate the number count of local maxima can be explained 
by the combination of the halo mass function and the correspondence between maxima and haloes.
Furthermore, the number count of local minima would 
bring additional information about the modification of gravity.

\vspace{2mm}
\item 
Throughout this paper, 
we consider the two models of $f(R)$ gravity
with the degree of freedom of a new scalar field 
$|f_{\rm R0}|=10^{-5}$ and $10^{-6}$.
We find that the stacked analysis around troughs 
and peak number counts can constrain on the model with
$|f_{\rm R0}|=10^{-5}$ with $\sim2\sigma$ level,
assumed ongoing imaging surveys with the sky coverage of a several
thousands squared degrees.
On the other hand, it is challenging 
to constrain the model with $|f_{\rm R0}|=10^{-6}$
with our proposed statistical methods.

\end{enumerate}

Although our findings would play an essential role 
to understand the nature of gravity with weak lensing measurement,
there are several caveats and limitations that 
must be kept in mind when interpreting the results presented in this work.

First, the simulation in this paper does not include any baryonic physics
and the possible effect due to the presence of massive neutrinos.
Both baryonic physics model and massive neutrinos
can affect the cosmic structure formation in various length scales.
\citep[e.g.,][]{
2010MNRAS.405.2161D,  2012MNRAS.423.2279C,
2014JCAP...12..053M, 2015JCAP...07..043C}.
In particular, our results on stacked signals around haloes 
(Section~\ref{sec.haloestack})
would potentially be affected by the baryonic physics at the inner regions,
while the massive neutrinos might have some impacts on the signals 
at the outskirts of each dark matter halo.
The simplest way to quantify these effects is to get together all effects
simultaneously: 
to run cosmological hydrodynamical simulation under the modified gravity
\citep[e.g.,][]{2013MNRAS.436..348P} 
in the presence of massive neutrinos.

There exist other statistical methods of weak lensing 
to extract cosmological information about the modification of gravity.
For instance, the two-point correlation of cosmic shear (or power spectrum)
has been proposed in previous works
\citep[e.g,][]{2007MNRAS.380.1029H}
and applied to existing data set \citep[e.g.,][]{2015MNRAS.454.2722H}.
Moreover, higher-order correlation function 
\citep[e.g,][]{2011JCAP...11..019G} and 
morphological statistics \citep{2015PhRvD..92f4024L} 
would be useful to improve the constraints on the nature of gravity
in weak lensing surveys.
Since our simulations can be helpful to 
understand the cosmological information content of 
weak lensing statistics, we plan to study them in details in future
(Shirasaki et al. in prep).

On the stacked analysis with voids, 
we should note that only $\sim10$ voids 
are found in each realization of our simulation.
In addition, we might underestimate the errors due to the limitation of simulation size.
Thus, larger weak lensing simulation would be 
more important to characterize the statistical property 
of cosmic voids and improve our understanding of the stacked signals
around voids under the modified gravity.
For this purpose, the full-sky simulation of weak gravitational lensing
\citep[e.g.,][]{2015MNRAS.453.3043S}
would be suitable to increase the sample size of voids and haloes.
A larger sample of voids and haloes would also 
enable us to study the mass assembly history in 
the modified gravity theory.

\section*{acknowledgments}
We thank an anonymous referee for careful reading and suggestion to improve the article. 
We would like to thank Baojiu Li for useful discussions and comments.
M.S. is supported by Research Fellowships of the Japan Society for the Promotion of Science (JSPS) for Young Scientists.
Numerical computations presented in this paper were in part carried out
on the general-purpose PC farm at Center for Computational Astrophysics,
CfCA, of National Astronomical Observatory of Japan.
Data analysis were [in part] carried out on common use data analysis computer system at the Astronomy Data Center, 
ADC, of the National Astronomical Observatory of Japan.

\bibliographystyle{mnras}
\bibliography{mn-jour,bibtex}
\end{document}